\documentclass[aps,prl,twocolumn,preprintnumbers,%superscriptaddress,
               showpacs,nofootinbib]{revtex4}
\newcommand{\PRE}[1]{}       % Use if journal style
\usepackage{bm}
\usepackage{epsfig}
\newcommand{\postscript}[2]{\setlength{\epsfxsize}{#2\hsize}
   \centerline{\epsfbox{#1}}}

\newcommand{\etal}{{\em et al.}}

\newcommand{\eqref}[1]{Eq.~(\ref{#1})}

\begin{document}

\preprint{CLNS 06/1964}

\title{
\PRE{\vspace*{1.5in}}
Big Bang Nucleosynthesis with Long-Lived Charged Massive Particles 
\PRE{\vspace*{0.3in}}
}

\author{Kazunori Kohri%
\PRE{\vspace*{.2in}}
}
\affiliation{Institute for Theory and Computation
Harvard-Smithsonian Center for Astrophysics, 
60 Garden Street, Cambridge, MA 02138, USA
\PRE{\vspace*{.1in}}
}

\author{Fumihiro Takayama%
\PRE{\vspace*{.2in}}
}
\affiliation{Institute for High Energy Phenomenology,
Cornell University, Ithaca, NY 14853, USA
\PRE{\vspace*{.1in}}
}

%\date{November 2002}

\begin{abstract}
\PRE{\vspace*{.1in}}

We consider Big Bang Nucleosynthesis (BBN) with  long-lived charged
massive particles. Before decaying, the long-lived charged particle  
recombines with a light element to form a bound state like a hydrogen atom.
This effect  modifies the nuclear reaction rates during the BBN epoch
through the modifications of the Coulomb field and the kinematics of
the captured light elements, which can change the light element
abundances. It is possible for heavier nuclei abundances
such as $^7$Li and $^7$Be to decrease sizably, while the ratios 
$Y_p$, D/H, and $^3$He/H remain unchanged. This may solve the
current discrepancy between the BBN prediction and the observed
abundance of $^7$Li. If future collider experiments find signals of
a long-lived charged particle inside the detector, the information of
its lifetime and decay properties could provide  insights into 
not only the particle physics models but also the phenomena
in the early Universe in turn.

\end{abstract}

%95.35.+d   Dark matter
%11.10.Kk   Field theories in dimensions other than four
%12.60.-i   Models beyond the standard model

\maketitle

%%%%%%%%%%%%%%%%%%%%%%%%%%%%%%%%%%%%%%%%%%%%%%%%%%%%%%%%%%%%%%%%%%%%%%
\section{Introduction}
%%%%%%%%%%%%%%%%%%%%%%%%%%%%%%%%%%%%%%%%%%%%%%%%%%%%%%%%%%%%%%%%%%%%%%

Recent cosmological observations agree remarkably with standard
$\Lambda$CDM models.  The one and three-year data of Wilkinson
Microwave Anisotropy Probe (WMAP) observation determined the
cosmological parameters to high precision~\cite{Spergel:2003cb,WMAP_HOME}.

In light of such recent progress of cosmological observations,
it has been shown that the Universe should be close to flat, and 
most of the matter must be in the form of non-baryonic dark matter, which
has been originally  considered as one of the best candidates to
explain an anomaly in the rotational curves of galaxies.

%Under considerations for the EW symmetry breaking sector in particle
%standard model and the extensions into high scale fundamental
%theories,  
In extension of the Standard Model explaining electroweak symmetry breaking 
and stability of the hierarchy, 
several candidates of the particle dark matter have been 
proposed such as the neutralino~\cite{Goldberg:1983nd},
the gravitino~\cite{superWIMP,GravitinoThermal,
GravitinoRecent1,GravitinoRecent2,GravitinoOld}, the axino~\cite{Axino} in
supersymmetric theory, branon dark matter~\cite{Cembranos:2003mr},
Kaluza Klein dark matter~\cite{Servant:2002aq,Cheng:2002ej}  and
Little Higgs dark matter~\cite{Birkedal:2006fz,Hubisz:2004ft} and so
on. The searches and the detailed studies of the dark matter have
become one of the most exciting aspects of near future collider experiments and
cosmological observations.

Considering such candidates in particle physics models, we expect
the large amount of the dark matter particle will be produced at the
near future colliders~\cite{Feng:2005gj}, which will be powerful tools 
to understand the properties of the dark matter ~\cite{DMcollider}.  
On the other hand, cosmological observations may provide information 
in new particle physics models, and even some implications on undetectable 
theoretical parameters in the collider experiments. Thus the connection of
cosmology to collider physics may provide wide possibilities to
understand the properties of the dark-matter particle and check the
cosmological models themselves.

At the present stage, the detailed properties of the dark matter is still
unknown. Therefore, even  exotic properties might be allowed. 
Future observations/experiments may prove them and single out or
constrain  dark-matter candidates. Even now, some problems in
cosmological observations may already show some hints to understand
the unknown properties of dark matter e.g., in the small scale
structure problem  ~\cite{cores,dwarfs,Navarro:1999fr} indicated in
the cold dark matter halo, the low $^7$Li problem~\cite{Cyburt:2003fe} 
and so on. There are several proposals to solve them by new physics
~\cite{DecayStruc,GraStruc,superWIMPStruc,CHAMPstruc,Wang:2004aq,
Li7superWIMP,Jedamzik:2004er,Spergel:1999mh,coll,WDM,Kaplinghat:2000vt,
Ichikawa:2004ju}. However, considerable astrophysical
uncertainties may still exist.
 
During the radiation dominated epoch well before the  decoupling of
the cosmic microwave background (CMB),  it is not necessary that the
dominant component of matter is neutral, and  that relic is the same
as the present one. For stable CHArged Massive Particles (CHAMPs)
~\cite{DeRujula:1989fe,Cahn:1980ss}, their fate in the universe had
been discussed~\cite{Boyd:1989},  and the searches for CHAMPs  inside
the  sea water were performed~\cite{CHAMPsearch}, which obtained null
results and got constraints on stable
CHAMPs~\cite{Kudo:2001ie}. According to their results, the production
of stable CHAMPs at future collider experiments is unlikely.
However, such null results can be applied only for the stable CHAMPs,
and  still the window for long-lived CHAMPs with a mass below O(TeV) 
is left open.  Such possibilities for the long-lived CHAMPs were
well-motivated in a scenario of super Weakly Interacting Massive Particle 
(superWIMP) dark matter~\cite{superWIMP}, which may inherit the desired 
relic density through the long-lived CHAMP decays.  The dominant component 
of the nonrelativistic (NR) matter during/after the BBN epoch might be charged
particles.  In supersymmetric theories, such a situation is naturally
realized  in gravitino lightest supersymmetric Particle (LSP) and axino 
LSP scenarios. Then the candidate for  the long-lived CHAMP would be a charged 
scalar lepton~\cite{GravitinoRecent1,GravitinoRecent2,Feng:2005ba}.

Trapping such long-lived CHAMPs,  the detailed studies of long-lived
charged particle will be possible in future collider experiments,
which may be able to provide  some nontrivial tests of underlying
theories, like measurement on the gravitino  spin, on the 
gravitational coupling in the gravitino LSP scenario 
~\cite{Buchmuller:2004rq}.  The trapping method in CERN LHC and International 
Linear Collider (ILC) has been performed in the context of supersymmetric 
theories~\cite{TrapCollider}. 
Also the collider phenomenology ~\cite{CHAMPcolliderOld,Feng:1997zr} and 
the other possible phenomena~\cite{CosmicStau,Hamaguchi:2006vp} have been 
discussed. 
%See also a
%recent related topic in Ref.~\cite{Hamaguchi:2006vp}.

In cosmological considerations of such long-lived particles,  the
effects on BBN by the late-time energy injection due to their decays
have been  studied in
detail~\cite{Kawasaki:2004yh,Kawasaki:2004qu,HadOld,EMold,Jedamzik:2006xz}.
On the other hand, in the past studies of the effects on the light
elements abundances, the analysis were simply applied to  long-lived
'charged' massive particles, assuming all CHAMPs are  ionized  and
freely propagating in the radiation dominated epoch well before the
CMB  decoupling.  However, we show that these results are not always
valid  if the bound state with a CHAMP and light elements may have
O(MeV)  binding energy~\cite{Cahn:1980ss}, and the bound state might
be stable against the destruction by the scattering off the huge
amount of the  background photons even during the BBN epoch. Also we
show that heavier elements tend to be captured at earlier time. Namely
the heavier light elements such as $^{7}$Li or $^{7}$Be form their
bound states earlier than the lighter light elements, D, T, $^3$He and
$^4$He.  Such a formation of the bound state with a heavy CHAMP may
provide possible changes of the nuclear-reaction rates and the
threshold energy of the reactions  and so on, which might result in the
change of the light element abundances.

What is the crucial difference from the case of electron captures?
In case of the electron capture, since the Bohr radius of an electron
is much larger than the typical pion-exchange length
O(1/$m_{\pi}$), two nuclei  feel the Coulomb barrier significantly
before they get close to each other.  On the other hand, in the case of
the capture of the  CHAMPs, the Bohr radius could be of the same order 
as the typical pion-exchange length. Then, the incident charged nuclei can
penetrate the weakened Coulomb barrier, and  the nuclear reaction
occurs relatively rapidly. The importance of such a bound state in 
the nuclear reaction had been identified for cosmic muons
~\cite{fusion,Ishida:2003}.
\footnote{In muon catalysis fusion, the formation of an atom containing 
two nuclei may be important.}

Concerning a discrepancy in $^7$Li between the standard big bang nucleosynthesis 
(SBBN) prediction by using the CMB baryon-to-photon ratio and the observational data, 
as we will show the details later, it is unlikely to attribute the
discrepancy only to uncertainties in nuclear-reaction rates in
SBBN~\cite{Serpico:2004gx,Cyburt:2004cq,Coc:2003ce}.  However as we
mentioned above,  if CHAMPs exist, the nuclear reaction rates during
the BBN epoch could be  changed from the  values known by experimental
data or observations of the sun, and may potentially solve the current
low $^7$Li/H problem.

If such long-lived CHAMPs existed and affected the light element
abundances, the lifetime would  be  long ($>$ 1sec). They may be
discovered as long-lived heavily ionizing  massive particles inside the
detector in the collider experiments. The measurements of their
lifetime and properties  may provide new insights  to understand not
only the particle physics models but also the phenomena in the early
Universe in turn.

In this paper, we discuss the possible change due to the long-lived 
CHAMPs  during/after BBN epoch and consider the effects on BBN.  
\footnote{In this paper, we use natural units for physical quantities.}

%%%%%%%%%%%%%%%%%%%%%%%%%%%%%%%%%%%%%%%%%%%%%%%%%%%%%%%%%%%%%%%%%%%%%%
\section{SBBN and observed light elements}
%%%%%%%%%%%%%%%%%%%%%%%%%%%%%%%%%%%%%%%%%%%%%%%%%%%%%%%%%%%%%%%%%%%%%%

The theory in SBBN has only one theoretical parameter, the baryon to
photon ratio $\eta$, to predict primordial light element
abundances. Comparing the theoretical predictions with observational
data, we can infer the value of $\eta$ in SBBN. It is well known that
this method had been the best evaluation to predict $\eta$ before WMAP
reported their first-year data of the CMB
anisotropy~\cite{Spergel:2003cb}.

\begin{figure}[tbp]
    \postscript{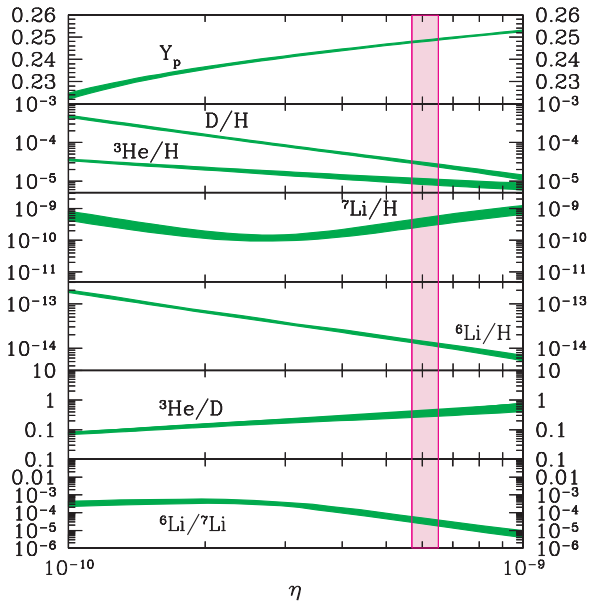}{0.95}
\caption{Theoretical predictions of 
$Y_p$, D/H, $^3$He/H, $^7$Li/H $^6$Li/H, $^3$He/D and $^6$Li/$^7$Li as
a function of the baryon-to-photon ratio $\eta$ in standard BBN with their
theoretical errors at
95 $\%$ C.L. The  WMAP value of $\eta$ at 95 $\%$ C.L.is also indicated 
as a vertical band. In the comparison between the BBN prediction and 
the central value of the observed abundances, it has been pointed out that 
the SBBN prediction with the WMAP value of the $\eta$ shows too high by a 
factor of a few in $^7$Li abundance and too low by several orders of magnitude 
in $^6$Li abundance if there is no late time $^6$Li production other than BBN.
~\cite{Cyburt:2003fe}
\label{fig:yield_sbbn}}
\end{figure}

WMAP observations have determined $\eta$ in high precision.  The value
of $\eta$ reported by the three-year WMAP
observations~\cite{WMAP_HOME} is
\begin{eqnarray}
\eta=\frac{n_b}{n_{\gamma}}=(6.10\pm 0.21)\times 10 ^{-10},
\end{eqnarray} 
where $n_b$ is number density of baryon, and $n_{\gamma}$ is number
density of the cosmic background photon. In Fig~\ref{fig:yield_sbbn} we
plot the theoretical prediction of the light element abundances with
their 2 $\sigma$ errors. The vertical band means the value of $\eta$
reported by the three-year WMAP observations at 2 $\sigma$.

We briefly discuss the current status of the theory of SBBN and the
observational light element abundances below, and check the
consistency with the CMB anisotropy observation. Further details
of the observational data are presented in a recent nice review by
G. Steigman~\cite{Steigman:2005uz}.  The errors of the following
observational values are at $1\sigma$ level unless otherwise
stated. Hereafter $n_X$ denotes the number density of a particle
$X$. $(X,C)$ denotes the bound state of CHAMP with an element $X$.

The primordial abundance of D  is inferred in the high redshift QSO
absorption systems. Recently  a new data was obtained at redshift
$z=2.525659$ toward Q1243+3074~\cite{Kirkman:2003uv}.  Combined with
these data~\cite{Tytler:1996eg,Burles:1998mk,O'Meara:2000dh,
Pettini:2001yu}, the primordial abundance is given as
${n_{\text{D}}}/{n_{\text{H}}}|_{\text{obs}}
=(2.78^{+0.44}_{-0.38})\times 10^{-5}$.~\footnote{Some of the observed
data have large dispersion than expected  and might have systematic
errors which may cause higher D/H~\cite{Kirkman:2003uv,Burles:1998mk}.}
%\begin{eqnarray}
%\frac{n_{\text{D}}}{n_{\text{H}}}|_{\text{obs}}
%=(2.78^{+0.44}_{-0.38})\times 10^{-5}.
%\end{eqnarray}
It agrees excellently with the value of $\eta$ predicted in the CMB
anisotropy observation. 

The abundance of $^3$He can increase and decrease through the chemical
evolution history. However, it is known that the fraction
$n_{^3\text{He}}/n_{D}$ is a monotonically increasing function of the
cosmic time~\cite{Sigl:1995kk,Kawasaki:2004qu}. Therefore the
presolar value is an upper bound on the primordial one,
$n_{^3\text{He}}/n_{D} < 0.59 \pm 0.54~(2 \sigma)$~\cite{Geiss93}. In
SBBN the theoretical prediction satisfies this constraint.

The primordial abundance of $^4$He is obtained from the recombination
lines  from the low-metallicity extragalactic HII region. The mass
fraction of the $^4$He  is inferred by taking the zero metallicity
limit as $\text{O/H}\to 0$ for the observational data~\cite{FieOLi}.
A recent analysis by Fields and Olive obtained the following  value by
taking into account the effect of the HeI absorption,
$Y(\text{FO})_{\text{obs}}=0.238\pm (0.002)_{\text{stat}} %%
\pm (0.005)_{\text{syst}}$,
%\begin{eqnarray}
%Y(\text{FO})_{\text{obs}}=0.238\pm (0.002)_{\text{stat}} \pm 
%(0.005)_{\text{syst}},
%\end{eqnarray}
where the first and second errors are the statistical and systematic ones.
 On the other hand, Izotov and Thuan~\cite{Izotov:2003xn} 
reported a slightly higher value,  $Y(\text{IT})_{\text{obs}}= %%
0.242\pm (0.002)_{\text{stat}}(\pm(0.005)_{\text{syst}})$
%\begin{eqnarray}
%Y(\text{IT})_{\text{obs}}=0.242\pm (0.002)_{\text{stat}}(\pm 
%(0.005)_{\text{syst}}),
%\end{eqnarray}
where we have added the systematic errors
following~\cite{Olive:1994fe, Olive:1996zu,Izotov:mi}. Olive and
Skillman recently reanalyzed the Izotov-Thaun data~\cite{Izotov:1998}
and obtained a much milder constraint~\cite{Olive:2004kq},
$Y(\text{OS})_{\text{obs}}=0.249\pm 0.009$.
%%
%\begin{eqnarray}
%    \label{eq:olive-skillman}
%    Y(\text{OS})_{\text{obs}}=0.249\pm 0.009
%\end{eqnarray}
%%
Even if we adopted the more restrictive value in Ref.~\cite{FieOLi}, SBBN is
consistent with CMB.

For $^7$Li, it is widely believed that the primordial abundance is
observed in  Pop II old halo stars with temperature higher than
$\sim$6000K and with low metallicity as a ``Spite's plateau''
value. The measurements by Bonifacio et al.~\cite{Bonifacio:2002yx}
gave $\text{Log}_{10}[n_{^7\text{Li}}/n_{\text{H}}]|_{\text{obs}}=
-9.66 \pm (0.056)_{\rm stat} \pm (0.06)_{\rm sys}$.
%$\frac{n_{^7\text{Li}}}{n_{\text{H}}}|_{\text{obs}}
%=(2.19^{+0.46}_{-0.38})\times 10^{-10}$.
%\begin{eqnarray}
%\frac{n_{^7\text{Li}}}{n_{\text{H}}}|_{\text{obs}}
%=(2.19^{+0.46}_{-0.38})\times 10^{-10}
%\end{eqnarray}
On the other hand, a significant dependence of $^7$Li on the Fe
abundance in the low metallicity region was reported in~\cite{RNB}.  If we
take a serious attitude towards this trend, and assume that this
comes  from the cosmic-ray interaction~\cite{RBOFN}, the  primordial
value is 
%$n_{^7\text{Li}}/n_{\text{H}}|_{\text{obs}}
%=(1.23^{+0.68}_{-0.32})\times 10^{-10}$
%%
\begin{eqnarray}
    \label{eq:ryan_li7}
    \frac{n_{^7\text{Li}}}{n_{\text{H}}}|_{\text{obs}}
=(1.23^{+0.32}_{-0.25})\times 10^{-10}~~~(at~~68\%~~\text{C.L}).
\end{eqnarray}
Even if  we adopt the higher value in Ref.~\cite{Bonifacio:2002yx},
the theoretical prediction is excluded at 2 $\sigma$ outside the
outskirts of observational and theoretical errors. Therefore when we
adopt the lower value in (\ref{eq:ryan_li7}), the
discrepancy worsens. The centeral value of the observation
is smaller than that of  SBBN  by a factor of about 3. This $^7$Li
problem has been pointed out by a lot of authors, e.g., see
Ref.~\cite{Cyburt:2003fe}.

It has been thought optimistically that this discrepancy would be
astrophysically  resolved by some unknown systematic errors in the
chemical evolution such as the uniform depletion in the convective
zone in the stars~\footnote{ See the recent report about spectroscopic
observations of stars in the metalpoor globular cluster NGC 6397 that
revealed trends of atmospheric abundance with evolutionary stage of
Lithium~\cite{Korn:2006tv}}. So far the researchers have added a large
systematic errors into the observational constraint by
hand~\cite{factor-of-two,Eidelman:2004wy}.

However, recently the plateau structure of $^{6}$Li in nine out of 24 Pop
II old halo stars was reported by Asplund et
al.~\cite{Asplund:2005yt}. The observed values of the isotope ratio
$n_{^{6}\text{Li}}/n_{^{7}\text{Li}}$ uniformly scatter between
$\simeq$ 0.01 and 0.09 at 2 $\sigma$, independently of the
metallicity, and are approximately similar to the previous
observational data ($=0.05 \pm 0.02$ at 2
$\sigma$~\cite{li6_obs}). Because the estimated $^{7}$Li abundance in
such stars is $n_{^7\text{Li}}/n_{\text{H}}|_{\text{obs}}=(1.1 - 1.5)
\times 10^{-10}$, the upper bound on the primordial $^{6}$Li agrees
with SBBN.  Although so far some models of the $^{6}\text{Li}$ and
$^{7}\text{Li}$ production through the cosmic-ray spallation of CNO
and $\alpha$-$\alpha$ inelastic scattering have been studied, the
predicted value of $n_{^{6}\text{Li}}/n_{^{7}\text{Li}}$ or
$n_{^{6}\text{Li}}/n_{\rm H}$ is obviously an increasing function of 
a metallicity~\cite{li6metal,li6LSTC,li6FO,suzuki:2002}.

As we have discussed, to be consistent with the SBBN prediction and
WMAP observations, we need a certain uniform depletion mechanism of
$^7$Li. Because $^{6}$Li is more fragile than $^{7}$Li, whenever
$^7$Li is destroyed in a star, $^{6}$Li suffers from the depletion,
too. If we require the primordial abundance of $^7$Li to be uniformly
depleted to a smaller value by a factor of three, the ratio
$^{6}$Li/$^7$Li might have to be reduced by a factor of ${\cal
O}$(10)~\cite{Pinsonneault:1998nf}. Therefore,  we do not have any
successful chemical evolution models at the present, to consistently
explain the observational value of $^{6}$Li/$^7$Li by starting from
the theoretical prediction of the primordial values of $^{6}$Li and
$^7$Li in the framework of SBBN.

Thus, by adopting the $\eta$ predicted in the CMB observations, we
would now have to  check SBBN itself or modified scenarios related with
BBN compared with the observational light element abundances.

%%%%%%%%%%%%%%%%%%%%%%%%%%%%%%%%%%%%%%%%%%%%%%%%%%%%%%%%%%%%%%%%%%%%%%
%%%%%%%%%%%%%%%%%%%%%%%%%%%%%%%%%%%%%%%%%%%%%%%%%%%%%%%%%%%%%%%%%%%%%%

In recent studies, it has been pointed out that  the uncertainties on
nuclear-reaction rates in SBBN never solve the  discrepancy of $^7$Li
between the theory and the observation. That is because the
uncertainties are highly constrained by known experimental data and
observations of the standard solar model.  In Ref.~\cite{Coc:2003ce},
the possible nuclear uncertainties were investigated.  It was shown
that only a nuclear reaction rate more than 100 times larger in
$^7$Be(n,$\alpha$)$^4$He and $^7$Be(d,p)2$^4$He might provide sizeable
change in the $^7$Li abundance. Notice that $^7$Be(n,$\alpha$)$^4$He does
not have an s-wave resonance due to the symmetry of the outgoing channel
while $^7$Be(n,p)$^7$Li has it. Since in the important energy region in 
the SBBN reaction $T\sim$ 50 keV, which is near threshold of the processes, 
the contribution to $^7$Be from $^7$Be(n,$\alpha$)$^4$He is negligible 
in SBBN relative to $^7$Be(n,p)$^7$Li because of the p-wave nature of the 
process. For $^7$Be(d,p)2$^4$He, the possibility  may not work in the light 
of the recent experimental data ~\cite{Angulo:2005mi}.  Also Cyburt
et.al.~\cite{Cyburt:2003ae} discussed the uncertainties on the
normalization of cross section for the process
$^3$He($\alpha$,$\gamma$)$^7$Be and  found that the uncertainties are
constrained in the light of a good agreement between the standard
solar model and solar neutrino data.

Therefore the remaining possibilities may be uncertainties on the
chemical evolution of Li from the BBN epoch to the present or effects
due to new physics.  Because now we do not have any successful
chemical evolution models, it must be important to consider the effect
of new physics.

As we mentioned before, the existence of CHAMPs might provide possible
change of nuclear-reaction rates during the BBN epoch, which may have
some impact on the prediction of primordial light element abundances.
In the next section, we will discuss the properties of the bound state
and the recombination  of CHAMP and the possible change of nuclear
reaction rates.

%%%%%%%%%%%%%%%%%%%%%%%%%%%%%%%%%%%%%%%%%%%%%%%%%%%%%%%%%%%%%%%%%%%%%%
\section{Bound state with a CHAMP and a light element}
%%%%%%%%%%%%%%%%%%%%%%%%%%%%%%%%%%%%%%%%%%%%%%%%%%%%%%%%%%%%%%%%%%%%%%

%%%%%%%%%%%%%%%%%%%%%%%%%%%%%%%%%%%%%%%%%%%%%%%%%%%%%%%%%%%%%%%%%%%%%%
\subsection{Evaluation of binding energy}
%%%%%%%%%%%%%%%%%%%%%%%%%%%%%%%%%%%%%%%%%%%%%%%%%%%%%%%%%%%%%%%%%%%%%%
We evaluate the binding energy for the bound state of a negatively
charged massive particle and a light element.  We simply consider the
case that the charged particle is a scalar.  The extension to a
fermion or the other higher spin cases would be  straightforward
although there  exist little differences.  Here we follow the way to
evaluate the binding energy assuming uniform charge distribution
inside the light element according to Ref.~\cite{Cahn:1980ss}.  Then
the Hamiltonian is represented by
\begin{eqnarray}
H=\frac{p^2}{2m_X}-\frac{Z_XZ_C\alpha}{2r_X}+\frac{Z_XZ_C\alpha}{2r_X}
(\frac{r}{r_X})^2,
\end{eqnarray}
for short distances $r < r_X$, and
\begin{eqnarray}
H=\frac{p^2}{2m_X}-\frac{Z_XZ_C\alpha}{r},
\end{eqnarray}
for long distances $r > r_X$, where $\alpha$ is the fine structure
constant, $r_X \sim$ 1.2 $A^{1/3}$/~200$~\text{MeV}^{-1}$ is the
nuclear radius, $Z_X$ is the electric charge of  the light element,
and $Z_C$ is the electric charge of the negatively charged massive particle.
$A$ is the atomic number, and  $m_X$ is the mass of the light element
$X$.  Here we assumed $m_X \ll m_C \sim {\cal O}$(100 GeV), which means
the reduced mass $1/\mu= 1/m_C+1/m_X\sim 1/m_X$.

For large nuclei, the exotic charged particle may be inside the
nuclear radius. The binding energy may be estimated under the harmonic
oscillator approximation by
\begin{eqnarray}
E_{\mbox{bin}}=\frac{3}{2}[\frac{Z_XZ_C\alpha}{r_X}-\frac{1}{r_X}
(\frac{Z_XZ_C\alpha}{m_X r_X})].
\end{eqnarray}
For small nuclei, the binding energy may be estimated well  as a
Coulomb bound state like a hydrogen atom,
\begin{eqnarray}
 E_{\mbox{bin}}\sim \frac{1}{2}{Z_X}^2{Z_C}^2\alpha^2m_X.
\end{eqnarray}
For intermediate regions in between the above cases, by using a trial
wave function,
%$\psi\sim e^{-\gamma r/r_X}$ {\bf with a constant $\gamma$}
we can express
\begin{eqnarray}
E_{\mbox{bin}}\sim \frac{1}{r_X}(\frac{1}{m_Xr_X}F(Z_XZ_C\alpha m_Xr_X)),
\end{eqnarray}
where $F(x)$ is  variationally determined~\cite{Cahn:1980ss}.  For
$0<Z_XZ_C\alpha m_Xr_X <1$, the Coulomb model gives a good
approximation. On the other hand, the harmonic oscillator
approximation gives  a better approximation for $2<Z_XZ_C\alpha m_Xr_X
< \infty$.

The binding energies are shown in Table~\ref{table:ebin}.
%%%%%%%%%%%%%%%%%%%%%%%%%%%%%%%%%%%%%%%%%%%%%%%%%%%%%%%%%%%%%%%%%%%%%%
\begin{table}
\begin{tabular}{|rc|rc|rc|} \hline
Nucleus(X) & & binding energy (MeV) & & atomic number & \\ \hline
$p$ & &0.025 & &Z=1 & \\
D & &0.050 & &Z=1 & \\
T & &0.075 & &Z=1 & \\
$^3$He & &0.270 & &Z=2 & \\
$^4$He & &0.311 & &Z=2 & \\ 
$^5$He & &0.431 & &Z=2 & \\
$^5$Li & &0.842 & &Z=3 & \\
$^6$Li & &0.914 & &Z=3 & \\
$^7$Li & &0.952 & &Z=3 & \\
$^7$Be & &1.490 & &Z=4 & \\
$^8$Be & &1.550 & &Z=4 & \\ 
$^{10}$B & &2.210 & &Z=5 & \\ \hline
\end{tabular}
\caption{Table of the binding energies for the various nuclei 
in the case of $Z_C=1$ given in Ref.~\cite{Cahn:1980ss}. 
For elements heavier than $^8$Be, 
the binding energies are given by the harmonic oscillator
approximation.}
\label{table:ebin}
\end{table}
%%%%%%%%%%%%%%%%%%%%%%%%%%%%%%%%%%%%%%%%%%%%%%%%%%%%%%%%%%%%%%%%%%%%%%
For a CHAMP with $Z_C=1$ and lighter elements ($p$, D, and T),
typically $Z_XZ_C\alpha m_Xr_X<1$. Thus the Coulomb approximation
works well.  However, for heavier elements such as Li or Be, there may
exist deviations which are more than ${\cal O}${(10)} percent.  For
elements lighter than $^8$B, the binding energy is still below  the
threshold energy of any nuclear reactions. If the atomic number is not
large like Li and Be, we can ignore the effects due to finite size
and the internal structure (excitations to higher levels and so on) as
a good approximation to calculate the capture cross section  and the
nuclear-reaction rates.

%%%%%%%%%%%%%%%%%%%%%%%%%%%%%%%%%%%%%%%%%%%%%%%%%%%%%%%%%%%%%%%%%%%%%%
\section{Capture of CHAMPs in the early universe}
%%%%%%%%%%%%%%%%%%%%%%%%%%%%%%%%%%%%%%%%%%%%%%%%%%%%%%%%%%%%%%%%%%%%%%

%%%%%%%%%%%%%%%%%%%%%%%%%%%%%%%%%%%%%%%%%%%%%%%%%%%%%%%%%%%%%%%%%%%%%%
\subsection{Recombination cross section}
%%%%%%%%%%%%%%%%%%%%%%%%%%%%%%%%%%%%%%%%%%%%%%%%%%%%%%%%%%%%%%%%%%%%%%
We evaluate the recombination cross section from the free state to the 1S
bound state assuming a hydrogen-type bound state  through a dipole
photon emission~\cite{Bethe:1957text}  and a pointlike particle for
the captured light element.  Then the cross section is
\begin{eqnarray}
\sigma_r v
&=&\frac{2^9\pi^2 \alpha Z_X^2}{3}
\frac{E_{\text{bin}}}{m_X^3v}
(\frac{E_{\text{bin}}}{E_{\text{bin}}+
\frac{1}{2}m_Xv^2})^2
\nonumber\\
&&~~~~~~~~~~~~~~~\times\frac{e^{-4\sqrt{\frac{2E_{\text{bin}}}{m_X v^2}}
\tan^{-1}(\sqrt{\frac{m_X v^2}{2E_{\text{bin}}})}}}
{1-e^{-2\pi\sqrt{\frac{2E_{\text{bin}}}{m_X v^2}}}}
\nonumber\\
&\simeq&\frac{2^9\pi^2 \alpha Z_X^2}{3e^4}
\frac{E_{\text{bin}}}{m_X^3v},
\end{eqnarray}
where $v$ is the relative velocity of a CHAMP and a light element.
Note that we have $m_Xv^2/2\simeq 3T/2\ll E_{\text{bin}}$ for
NR particles in kinetic equilibrium.  Here we use
the Coulomb model (hydrogen type) to evaluate the capture
rate~\cite{DeRujula:1989fe}, where the binding energy
$E_{\text{bin}}=\alpha^2 Z_C^2Z_X^2m_X/2$ and the Bohr radius
$r_B^{-1}\simeq \alpha Z_CZ_Xm_X$.

The thermal-averaged cross section is written as
\begin{eqnarray}
&&\langle \sigma_r v \rangle = \frac{1}{n_1n_2}
(\frac{g}{(2\pi)^3})^2\int d^3p_1 d^3p_2 e^{-\frac{(E_1+E_2)}{T}} 
\sigma_r v\nonumber\\
&=&\frac{1}{n_Gn_r}(\frac{g}{(2\pi)^3})^2
\int d^3p_G e^{-\frac{m_G}{T}} e^{-\frac{p_G^2}{2m_GT}}
\int d^3 p_r \sigma_{r} v e^{-\frac{p_r^2}{2\mu T}} \nonumber\\
&=&\frac{2^9\pi \alpha Z_X^2\sqrt{2\pi}}{3e^4}
\frac{E_{\text{bin}}}{m_X^2\sqrt{m_XT}},
\end{eqnarray}
where $m_G=m_1+m_2$ and $\mu=m_Xm_C/(m_X+m_C)\simeq m_X$ with
\begin{eqnarray}
&&n_G=\frac{g}{(2\pi)^3}\int d^3p_G e^{-\frac{m_G}{T}} 
e^{-\frac{p_G^2}{2m_GT}}\nonumber,\\
&&n_r=\frac{g}{(2\pi)^3}\int d^3p_r e^{-\frac{p_r^2}{2\mu T}}\nonumber.
%\\
%&&\frac{1}{n_r}\frac{g}{(2\pi)^3}\int d^3 p_r 
%\frac{\mu}{p_r} e^{-\frac{p_r^2}{2\mu T}}
%=\sqrt{\frac{2\mu}{\pi T}}.
\end{eqnarray}
Here we have assumed that only one CHAMP is captured by a nucleus.
Since the photon emission from a CHAMP is suppressed, the
recombination cross section for the further capture of an additional
CHAMP by the bound state would be much smaller. Therefore, as a first
step, it would be reasonable to ignore the multiple capture of CHAMPs
by a nucleus.

Here we have estimated only the direct transition from the free state
into  the 1S bound state.  However, if the transition from higher
levels into the 1S state is sufficiently rapid against the destruction due
to scatterings off the thermal photons, even the capture into the
higher levels might contribute to the recombination of a CHAMP. The
typical time scale of the transition from $n$th level into the 1S state
is $1/(E_{\text{bin}:1S}-E_{\text{bin}:n}) \sim
O(1/E_{\text{bin}:1S})$ where $E_{\text{bin}:n}$ is  the binding
energy of the  $n$th level. Up to some levels,  this time scale might
be shorter than  the destruction rate after the 1S state became
stable.  However, such higher-level captures would not significantly
enhance the recombination  cross section because the capture rate into
higher levels is relatively suppressed and small.

For highly charged massive nuclei or elements heavierthan boron, the
binding energies with CHAMPs can become of the order of magnitude of the
excitation energies of nucleons inside the nuclei, or even of the same order
of  magnitude of the nuclear binding energies. In such cases, the
capture process of light elements by CHAMPs may be nontrivial. In
addition, to correctly calculate the capture rates, we would have to
understand the modification by the effects due to not only the finite
size but also the internal structure of the light element. In this
paper, we ignore these effects because they
are unimportant since we consider lighter nuclei up to Li and Be.

%%%%%%%%%%%%%%%%%%%%%%%%%%%%%%%%%%%%%%%%%%%%%%%%%%%%%%%%%%%%%%%%%%%%%%
\subsection{Case in kinetic and chemical equilibrium}
%%%%%%%%%%%%%%%%%%%%%%%%%%%%%%%%%%%%%%%%%%%%%%%%%%%%%%%%%%%%%%%%%%%%%%
To evaluate the number density of the captured CHAMPs,  we would be
able to use the thermal relation among chemical potentials if the
capture reactions well establish the chemical equilibrium between the
CHAMPs and  the light elements.  The number density is determined by
the following Saha equation,
\begin{eqnarray}
n_{(X,C)}=\frac{2}{\pi^2}
\zeta(3) \frac{n_X}{n_{\gamma}} n_C(\frac{2 \pi T}{m_X})^{3/2}
e^{\frac{E_{\text{bin}}}{T}}
\end{eqnarray}
where $n_X$ and $n_{\gamma}$ are number densities of a light element $X$
 and thermal photons, and $E_{\text{bin}}$ is the binding energy of 
the light element. 

%%%%%%%%%%%%%%%%%%%%%%%%%%%%%%%%%%%%%%%%%%%%%%%%%%%%%%%%%%%%%%%%%%%%%%
\subsection{General cases}
%%%%%%%%%%%%%%%%%%%%%%%%%%%%%%%%%%%%%%%%%%%%%%%%%%%%%%%%%%%%%%%%%%%%%%
However the question of whether such kinetic and chemical equilibrium are
well established among all light elements and CHAMP is nontrivial.  
Here we consider the Boltzmann equations for CHAMPs, a light element $X$  
and the bound state  $(X,C)$. For CHAMPs,
\begin{eqnarray}
 \frac{\partial}{\partial t}n_{C} +3Hn_{C}
= \left[\frac{\partial}{\partial t}n_{C}\right]_{\mbox{capture}},
\end{eqnarray}
where $H$ is the Hubble expansion rate. For a light element $X$,
\begin{eqnarray}
\frac{\partial}{\partial t}n_{X}
+3Hn_{X}=
\left[\frac{\partial}{\partial t}n_{X}\right]_{\mbox{fusion}}
+\left[\frac{\partial}{\partial t}n_{X}\right]_{\mbox{capture}}.
\end{eqnarray}
For the bound state,
\begin{eqnarray}
&\lefteqn{ \frac{\partial}{\partial t}n_{(C,X)}  
+3Hn_{(C,X) } }     \nonumber \\
&=& \left[ \frac{\partial}{\partial t}n_{(C,X)} \right]_{\mbox{fusion}} 
- \left[\frac{\partial}{\partial t}n_{X} \right]_{\mbox{capture}}.
\end{eqnarray}

By using the detailed balance relation between the forward process
 $X+C\to\gamma+(X,C)$ and the reverse process 
$(X,C)+\gamma\to X+C$, the capture reaction may be written by
\begin{eqnarray}
\lefteqn{ \left[ \frac{\partial}{\partial t}n_{X}
\right]_{\mbox{capture}} = 
\left[ \frac{\partial}{\partial t}n_{C}
\right]_{\mbox{capture}}
} 
\nonumber \\
&\simeq& -\langle \sigma_r v \rangle \left[
  n_Cn_X-n_{(C,X)}n_{\gamma}(E>E_{\text{bin}})\right], 
\end{eqnarray}
where
\begin{eqnarray}
&& n_{\gamma}(E>E_{\text{bin}})\equiv n_{\gamma}
\frac{\pi^2}{2\zeta(3)}(\frac{m_X}{2\pi T})^{3/2}
e^{-\frac{E_{\text{bin}}}{T}},
\end{eqnarray}
and 
\begin{eqnarray}
&& n_{\gamma}=\frac{2\zeta(3)}{\pi^2}T^3.
\end{eqnarray}
For a light element, if $\langle \sigma_r v \rangle n_C/H  \gg 1$ is
satisfied and the kinetic equilibrium is well established,  we can get
the Saha equation by requiring an equilibrium condition
$[\frac{\partial}{\partial t}n_X]_{\mbox{capture}}=0$ in this
equation.  Since we are interested in the time evolution of not only
CHAMPs but also light elements, we carefully study the case of
$\langle \sigma_r v \rangle n_C/H > 1$ even in the  case of $ \langle
\sigma_r v \rangle n_X/H\ll 1$.

%%%%%%%%%%%%%%%%%%%%%%%%%%%%%%%%%%%%%%%%%%%%%%%%%%%%%%%%%%%%%%%%%%%%%%
\subsection{Critical temperature at which a bound state is formed}
%%%%%%%%%%%%%%%%%%%%%%%%%%%%%%%%%%%%%%%%%%%%%%%%%%%%%%%%%%%%%%%%%%%%%%
When the temperature is higher than the binding energy of light
elements, the destruction rate of bound states  by  scatterings off
the thermal photons  with $E>E_{\text{bin}}$ is rapid. Then only a
small fraction of bound states   can  be formed, $n_{(C,X)}\sim
n_Cn_X/n_{\gamma}(E>E_{\text{bin}})\ll n_X$. Once the temperature
becomes lower than the binding energy, the capture starts, and
the bound state becomes stable if the other destruction processes
among the nuclei  are inefficient.  \footnote{Note that the abundances
of heavier elements such as Li and Be are smaller  than those of
lighter elements ($p$, D, T and He). As we will see later, considering
the relic density of  relevant candidates of CHAMPs, their capture can
only affect on the abundance of the heavier elements. Our scenarios
would not significantly change the lighter element abundances.}  The
critical temperature at which the capture becomes efficient is
estimated  as follows.  In the case of $n_X>n_C$, taking $n_C\sim
n_{(C,X)}$, we get a relation,
\begin{eqnarray}
(\frac{m_X}{T})^{3/2} e^{-\frac{E_{\text{bin}}}{T}}\sim 
\frac{n_X}{n_{\gamma}}= {\cal O}(10^{-10}).
\end{eqnarray}
On the other hand, in the case of $n_X<n_C$, taking $n_X\sim
n_{(C,X)}$, we have
\begin{eqnarray}
&&\lefteqn{(\frac{m_X}{T})^{3/2}  e^{-\frac{E_{\rm bin}}{T}} }
 \nonumber \\ && \sim
 \frac{n_C}{n_{\gamma}}  
 \sim {\cal O}(10^{-10}) (\frac{100{\rm
 GeV}}{m_C})(\frac{\Omega_C}{0.23}).
\end{eqnarray}
This analysis shows that the critical temperature is approximately
\begin{eqnarray}
 T_c \simeq \frac{E_{\text{bin}}}{40}.
\end{eqnarray}
In case of $Z_C=1$, we find $T_{c}\sim E_{\text{bin}}/40 \sim 8$keV
for $^4$He.  

Here we consider the temperature where some fraction of $X$ is
captured by CHAMPs. For example, taking $n_{(C,X)}/n_X\simeq 10^{-5}$,
we get
\begin{eqnarray}
(\frac{m_X}{T})^{3/2} e^{-\frac{E_{\text{bin}}}{T}}\sim 
\frac{n_C}{n_{\gamma}}\frac{n_X}{n_{(C,X)}}=O(10^{-6}).
\end{eqnarray}
This condition is satisfied at $T_c^{(2)}\sim E_{\text{bin}}/30$. 
Since the abundance of $^4$He is large  below 0.1MeV,
even though  the only small fraction of $^4$He is trapped by CHAMPs,
there might be relevant effects caused by the captures.  
%However, we found that the SBBN processes are already decoupled 
%after $T< T_{c}$,and the changes from SBBN is negligible.

For protons, the efficient captures start at a temperature lower than
1 keV (at cosmic time longer than $10^{6}$ sec). Since the bound state
is neutral for a single-charged CHAMPs $Z_C=1$, and might be
negatively charged for a multi-charged CHAMPs $Z_C>1$, there is no
Coulomb repulsion anymore.  Thus, even the bound state can collide
with each other.  If the number density of CHAMPs is not too small, and
most CHAMPs are captured by protons, the change could be sizable
for longer lived of CHAMPs ($\tau > 10^6$ sec).
\footnote{Since the CHAMPs with a long lifetime more than $\gg
10^6$sec may induce the other effects on cosmology~\cite{CHAMPstruc}.}

%%%%%%%%%%%%%%%%%%%%%%%%%%%%%%%%%%%%%%%%%%%%%%%%%%%%%%%%%%%%%%%%%%%%%%
\subsection{Capture rate}
%%%%%%%%%%%%%%%%%%%%%%%%%%%%%%%%%%%%%%%%%%%%%%%%%%%%%%%%%%%%%%%%%%%%%%
Since the capture process competes with the expansion of the universe,
we have to check if the following relation holds during the
meaningful time, which ensures that the capture by CHAMPs is efficient
compared to the expansion rate of the universe,
\begin{eqnarray}
H \ll \langle  \sigma_r v \rangle n_C.
\end{eqnarray}
That is, the capture rate of a light element is controlled by the
following  $\kappa$,
\begin{eqnarray}
\lefteqn{\kappa\equiv\frac{\langle \sigma_r v \rangle n_C}{H}}\\
&=&2.6
\sqrt{\frac{3.2}{g_{\ast}}}\sqrt{\frac{T}{24\text{keV}}}
\left( \frac{Z_X}{3}\right )^4 \left( \frac{7\text{GeV}}{m_X} \right)^{3/2}
\frac{\Omega_C}{0.23}\frac{100\text{GeV}}{m_C}. \nonumber
\end{eqnarray}
$\kappa$ is approximately 2.6 and 0.43 for $^7$Li and $^4$He at their
critical  temperatures, respectively. Here we assumed that
$\Omega_C\simeq 0.23$ and $m_C=100$GeV.

In the evaluation of the capture rates for light elements,  we
considered  relatively large  number densities of the CHAMPs, which are
approximately similar to that of $^4$He or even more because here we
assumed that a CHAMP can decay into much lighter dark matter or
almost massless SM particles later.   Under these circumstances, we
naturally  expect a larger value of the capture rates than  the upper
limit in case of the stable CHAMP scenario. Of course, we have to
check that the decay never disturbs the successful concordance of cold 
dark matter (CDM) with large scale structure formation in  the Universe 
and so on. Later we will discuss this problem.

Next let us estimate the time evolution of $X$ itself and the capture
fraction of $X$ by a CHAMP.  At below the critical temperature $T_c$,
the destruction term of $(X,C)$ becomes negligible due to the Boltzmann
suppression.~\footnote{The ignorance of the destruction term at $T_c$ may 
be valid if the recommbination cross section is not too large. 
If the cross section is enough large, the number density of bound state 
may be well descriobed by the Saha equation.} 
Then the number densities of the light element $X$ and the
bound state of $X$ with a CHAMP, $(X,C)$ are obtained by solving the
following equations. Here  any destruction reactions
of $X$ would be negligible close to the end of the BBN epoch
($\lesssim$~50 keV.),
\begin{eqnarray}
&&\frac{d}{dT}(\frac{\eta_X}{\eta_X(T_c)})\simeq
\frac{\langle \sigma v \rangle n_C}{HT}\frac{\eta_X}{\eta_X(T_c)}, 
\label{eq:bol1}
\\
&&\frac{d}{dT}(\frac{\eta_{(C,X)}}{\eta_X(T_c)})=
-\frac{\langle \sigma_{r} v \rangle n_C}{HT}\frac{\eta_X}{\eta_X(T_c)}+
\text{(fusion~~part)}, \nonumber
\label{eq:bol2}
\end{eqnarray}
where $\eta_i=n_i/s$, and $\eta_X(T_c)$ is the initial number density
per entropy density when the capture starts,  assuming that the
standard processes of the light elements are (almost) frozen out. We
also assumed $n_C\gg n_X$ which is correct except for $^4$He. We find
that if $\kappa$ is larger than unity at the critical temperature, the
capture will be efficient.

\begin{figure}[tbp]
\postscript{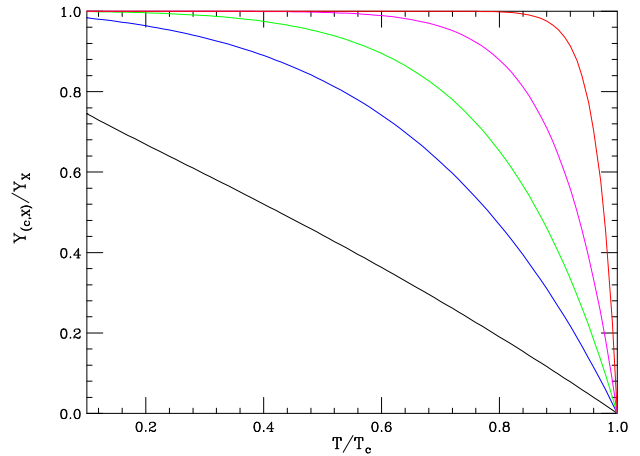}{0.95}
\caption{$\eta_{(C,X)}/\eta^0_X$ as a function of $T/T_c$ for 
$ \langle \sigma_{r} v \rangle n_C/H|_{T=T_c}$=1, 3, 5, 10, and 30
from left to right, respectively. Here we have ignored the standard
BBN processes. Also we have taken the initial condition as $\eta_X(T_c)\sim 0$.
\label{fig:plot2} If $ \langle \sigma_{r} v \rangle n_C/H|_{T=T_c}\gg 1$, 
the Saha equation will be a good approximation and the capture will 
immidiately occur at $T\sim T_c$. On the other hand, if 
$ \langle \sigma_{r} v \rangle n_C/H|_{T=T_c}\sim 1$ or less than 1, 
the approximation by the Saha equation may be falied.}
\end{figure}

Ignoring the fusion part of the standard processes in
Eq.(\ref{eq:bol1}),  we find the following analytical solution of
$\eta_X(T)$,
\begin{eqnarray}
\eta_X(T)=\eta_X(T_c)e^{-2\kappa_{i} (1-\sqrt{T/T_c})},
\end{eqnarray}
where $\kappa_i= \langle \sigma_{r} v \rangle n_C/H|_{T=T_c}$.
\footnote{For $^7$Be, $^7$Li and lighter elements, above approximation
works well if $Z_c$ is close to 1. As we will see later, the change
of nuclear  reaction rates does not modify the fusion part of the
noncaptured light elements so much because the most of reverse
processes has already been decoupled even after the other elements are
captured by CHAMPs.} For the numerical solution of
Eq.~(\ref{eq:bol2}), see Fig.~\ref{fig:plot2}.

For a more precise analysis, especially for the  Boltzmann equation of
the CHAMP bound state, we may have to take into account  the nuclear
reaction processes simultaneously.  Therefore we will need to do the
numerical calculations to solve the Boltzmann equations  including
both the capture and the BBN processes in the future~\cite{takayama:2006}.
However, to qualitatively understand how large changes would be
possible, for simplicity we assume only the instantaneous captures
in the current work.

%%%%%%%%%%%%%%%%%%%%%%%%%%%%%%%%%%%%%%%%%%%%%%%%%%%%%%%%%%%%%%%%%%%%%%
%%%%%%%%%%%%%%%%%%%%%%%%%%%%%%%%%%%%%%%%%%%%%%%%%%%%%%%%%%%%%%%%%%%%%%

%%%%%%%%%%%%%%%%%%%%%%%%%%%%%%%%%%%%%%%%%%%%%%%%%%%%%%%%%%%%%%%%%%%%%%
\section{Change of nuclear reaction rates in BBN by the capture of CHAMPs}
%%%%%%%%%%%%%%%%%%%%%%%%%%%%%%%%%%%%%%%%%%%%%%%%%%%%%%%%%%%%%%%%%%%%%%
The capture of light elements by CHAMPs weakens the Coulomb
barrier in the nuclear reactions during/after the BBN epoch.  The
change of nuclear reaction rates could become large because the
Coulomb factor exponentially suppresses the reaction rates. In
general,  the reaction rates among charged nuclei during the BBN epoch
are determined by the competition between the coulomb suppression and
the Boltzmann suppression,  which play important roles to determine
the freeze-out of light  element abundances at the end of the BBN
epoch.  Considering the corrections on these two exponential
suppressions, we will next consider the possible changes of nuclear
reaction rates.

%%%%%%%%%%%%%%%%%%%%%%%%%%%%%%%%%%%%%%%%%%%%%%%%%%%%%%%%%%%%%%%%%%%%%%
\subsection{Coulomb potential and scattering problem}
%%%%%%%%%%%%%%%%%%%%%%%%%%%%%%%%%%%%%%%%%%%%%%%%%%%%%%%%%%%%%%%%%%%%%%
If there are Coulomb expulsion forces, the wave function of an
incident particle would be exponentially suppressed at the target.
Since we use a plane wave for the wave function  to evaluate the
incident flux at a sufficiently far place from the target,  the real
flux which is associated with the reaction  would be evaluated by
renormalizing the wave function.  Since the change of the
wave-function normalization from the plane wave  is associated with the
state before the nuclear reaction,  it is independent of the short
distance nuclear reaction by nuclei.  We can expect that the Coulomb
factor is factorized as follows
~\footnote{This factorization may be valid only if the Bohr radius of 
bound state is not too large relative to the radius of the bound and 
incident nuclei. If the Bohr radius is large, which may be expected in 
$Z_X=1$ nuclei cases, we have to understand how the bound state is 
disturbed by the incident nucleus. In such large Bohr radius cases, 
for example, to proceed nuclear fusion, the hydrogen-type bound state of 
the nucleus and a CHAMP may have to constitute a molecule before the nuclear fusion. 
Then we will have to evaluate the capture reaction rate of molecule.},
\begin{eqnarray}
F_{ab}(v)=\frac{2\pi Z_aZ_b\alpha/v}
{e^{\frac{2\pi Z_aZ_b\alpha}{v}}-1}
\simeq \frac{2\pi Z_aZ_b\alpha}{v} 
e^{-\frac{2\pi Z_aZ_b\alpha}{v}}.
\end{eqnarray}

After a CHAMP  is trapped by a light element  $a$,  for a collision
between a bound state (CHAMP+ the light element $a$)  and a light
element $b$,
\begin{eqnarray}
F_{(aC)b}(\beta)
\simeq \frac{2\pi \alpha Z_{(aC)}Z_b}{\beta} 
e^{-\frac{2\pi \alpha Z_{(aC)}Z_b}{\beta} },
\end{eqnarray}
where $Z_{(aC)}=Z_a-Z_C$. Note that  $\beta$ is the relative velocity
between the bound state $(aC)$ and  the $b$ element, not the $a$ and
the $b$ element.  Hence $\beta$ could be slightly  different from $v$
which is the normal relative velocity between the thermal $a$ and  the
thermal $b$.  Here we assumed that a light element can capture only
one CHAMP (with the charge $Z_C$).  

For the case of nuclear reactions through a collision between
charged bound sates (CHAMP + light element $a$ and CHAMP + light element $b$), 
the Coulomb-penetration ability is determined by the relative 
velocity  between the bound states. That is,
\begin{eqnarray}
F_{(aC)(bC)}(\beta_2)
=\frac{2\pi Z_{(aC)}Z_{(bC)}\alpha}{\beta_2} 
e^{-\frac{2\pi Z_{(aC)}Z_{(bC)}\alpha}{\beta_2}},
\end{eqnarray}
where $\beta_2=p_r/\mu_{(aC)(bC)}\simeq O(T/m_C)<< \beta=O(T/m_X)$.
Under these circumstances, the collision between charged bound states may
be highly suppressed relative to the standard BBN reactions because
$Z_{(aC)}Z_{(bC)}/\beta_2 > Z_a Z_b /v$ if $Z_X > Z_C$.  This bound
state - bound state collision might become important if a huge
number of CHAMPs are captured by $^4$He.  However,   the
typical temperature to start capture is below {\cal O}(10) keV, and
the Coulomb factors for the normal nuclear reactions in SBBN is
highly suppressed, and have already been decoupled
by that time. Thus this type of collision will not contribute
to any sizable changes of the light element abundances

For $Z_X=1$ ($Z_C=1$) cases like protons, since there is no Coulomb 
suppression because the bound state is neutral, the collision between two 
bound states may be important.

%%%%%%%%%%%%%%%%%%%%%%%%%%%%%%%%%%%%%%%%%%%%%%%%%%%%%%%%%%%%%%%%%%%%%%
\subsection{SBBN and thermal-averaged fusion rates}
%%%%%%%%%%%%%%%%%%%%%%%%%%%%%%%%%%%%%%%%%%%%%%%%%%%%%%%%%%%%%%%%%%%%%%
First, we discuss nuclear-reaction rates in SBBN, and next  we will
extend the discussions to the cases with the CHAMPs.

For simplicity, we consider the case of $2 \to 2$  non-resonant
reactions among charged light nuclei.  The other cases may be
straightforward through similar discussions.  In a  SBBN  process
$a+b\to c+d$, the forward process and the reverse process are defined
by the difference between the total masses in the initial and  the
final state. If ${\it Q}_{ab,cd}=m_a+m_b-m_c-m_d={\it Q}_{\text{SBBN}}>0$,  the
process $a+b\to c+d$ has no threshold and is called the forward
process. On the other hand, the process $c+d\to a+b$ has threshold
({\it Q}-value ${\it Q}_{cd,ab}=-{\it Q}_{\text{SBBN}}<0$) and is  called the reverse
process of $a+b\to c+d$ .  Usually the reverse process $c+d\to a+b$
has a strong Boltzmann suppression  by $e^{-{\it Q}_{\text{SBBN}}/T}$ if
the {\it Q} value is larger than the Gamow peak energy  of the process.

%%%%%%%%%%%%%%%%%%%%%%%%%%%%%%%%%%%%%%%%%%%%%%%%%%%%%%%%%%%%%%%%%%%%%%
\subsubsection{SBBN reaction rates with no threshold}
%%%%%%%%%%%%%%%%%%%%%%%%%%%%%%%%%%%%%%%%%%%%%%%%%%%%%%%%%%%%%%%%%%%%%%
Naively the nuclear reactions of SBBN occur at almost the threshold region. 
Thus the cross section may be well described by the 
lower partial wave modes. Taking into account for the discussion of the 
wave function normalization in previous section, 
the reaction cross section is written as follow.
\begin{eqnarray}
\sigma_{\text{fusion}}v&=&(\sigma_S+\sigma_P v^2+....)F_{ab}(v)\nonumber\\
&=&\sigma_0v(v) \frac{2\pi Z_aZ_b\alpha}{v} 
e^{-\frac{2\pi Z_aZ_b\alpha}{v}}
\end{eqnarray}
where $\sigma_0v(v)=\sigma_S+\sigma_P v^2+....$.

Here we introduce a new variable, the ``astrophysical {\it S} factor'' which
astrophysicists have used in the calculation of nucleosynthesis,
\begin{eqnarray}
S(E_r)&=&\sigma_{\text{fusion}}E_r e^{\sqrt{\frac{E_G}{E_r}}}\nonumber\\
&=& \sigma_0v(v)  \pi Z_aZ_b\alpha\mu_{ab}
\end{eqnarray}
where $E_G=2\pi^2Z_a^2Z_b^2\alpha^2\mu_{ab}$ and
$E_r=p_r^2/2\mu_{ab}=\mu_{ab}  v^2/2$.  Notice that this {\it S} factor is a
function of the center-of-mass (CM) energy  and is inferred by the
measurements of $\sigma_{\text{fusion}}v$ in experiments and
observations. The recent fitting functions are given in
Refs.~\cite{Serpico:2004gx,Cyburt:2004cq}.

By using this {\it S} factor, we calculate the thermal-averaged cross
section.
\begin{eqnarray}
&&\langle \sigma_{\text{fusion}} v \rangle =
\frac{g}{(2\pi)^3n_r}\int d^3 p_r \sigma_{\text{fusion}} v 
e^{-\frac{p_r^2}{2\mu_{ab} T}}\nonumber\\
&=&\frac{8\pi g T \mu_{ab}}{(2\pi)^3n_r}\int d x S(xT) 
e^{-(x+\sqrt{\frac{x_G}{x}})}\nonumber\\
&=&\frac{8\pi gT\mu_{ab}}{(2\pi)^3n_r}\int d x S(xT) 
e^{-(\frac{3}{4^{1/3}}x_G^{1/3}+\frac{3}{4x_0}(x-x_0)^2+....)}
\nonumber\\
&\simeq& \frac{8\pi gT\mu_{ab}}{(2\pi)^3n_r} \sqrt{\frac{\pi x_0}{3}}
(1+Erfc(\frac{\sqrt{3x_0}}{2}))S(x_0 T) 
e^{-\frac{3}{4^{1/3}}x_G^{1/3}}\nonumber\\
&\simeq& \frac{8\pi gT\mu_{ab}}{(2\pi)^3n_r} \sqrt{\frac{4\pi x_0}{3}}
S(x_0 T) 
e^{-\frac{3}{4^{1/3}}x_G^{1/3}}
\end{eqnarray}
where $x=E_r/T$, $x_G=E_G/T$, and $x_0=(x_G/4)^{1/3}$. 
Since the main contribution of this integral comes from 
the stationary point of the exponent, we expanded the exponent around
 the stationary point $x_0=(x_G/4)^{1/3}$.

Finally we can evaluate the thermal-averaged nuclear-reaction rate among 
charged light elements.
\begin{eqnarray}
\langle \sigma_{\text{fusion}} v \rangle (T)
=\sqrt{\frac{32}{4^{1/3}}\frac{E_G^{1/3}}{3\mu_{ab}}}\frac{S(x_0 T)}{T^{2/3}}
e^{-\frac{3}{4^{1/3}}(\frac{E_G}{T})^{1/3}},
\end{eqnarray}
where $1/\mu_{ab}=1/m_a+1/m_b$.

%%%%%%%%%%%%%%%%%%%%%%%%%%%%%%%%%%%%%%%%%%%%%%%%%%%%%%%%%%%%%%%%%%%%%%
\subsubsection{SBBN reaction rates with threshold}
%%%%%%%%%%%%%%%%%%%%%%%%%%%%%%%%%%%%%%%%%%%%%%%%%%%%%%%%%%%%%%%%%%%%%%
We often evaluate reverse reaction rates from the experimental data of
forward reaction rates by using the detailed balance relation. For
example, in a $2\to 2$ non-resonant reaction $a+b\to c+d$,
\begin{eqnarray}
\frac{ \langle \sigma_{\text{fusion}} v \rangle
_{cd}}{\langle \sigma_{\text{fusion}} v \rangle_{ab}}= 
(\frac{\mu_{ab}}{\mu_{cd}})^{3/2}(\frac{m_a+m_b}{m_c+m_d})^{3/2}
\frac{g_ag_b}{g_cg_d}e^{-\frac{{\it Q}}{T}}
\end{eqnarray}
where {\it Q} is the {\it Q} value of the forward reaction and $g_a$ is the
number of degrees of freedom of the light element $a$. Notice that the factor
$e^{-{\it Q}/T}$  arises from the Boltzmann suppression for the high energy
component  with $E_r>{\it Q}$ in thermal distribution.

%%%%%%%%%%%%%%%%%%%%%%%%%%%%%%%%%%%%%%%%%%%%%%%%%%%%%%%%%%%%%%%%%%%%%%
\subsection{Extension to BBN with the captured CHAMP}
%%%%%%%%%%%%%%%%%%%%%%%%%%%%%%%%%%%%%%%%%%%%%%%%%%%%%%%%%%%%%%%%%%%%%%
We have shown that the collision among charged CHAMP bound states will not
result in any changes to  SBBN.  Here we focus on the nuclear-
reaction rate for the collision between  a bound state (CHAMP+light
element) and an unbound light element.~\footnote{For the case of
scatterings among neutral bound states,  the collision can easily
occur. In such cases the calculation is  straightforward.}

%%%%%%%%%%%%%%%%%%%%%%%%%%%%%%%%%%%%%%%%%%%%%%%%%%%%%%%%%%%%%%%%%%%%%%
\subsubsection{Forward and Backward process}
%%%%%%%%%%%%%%%%%%%%%%%%%%%%%%%%%%%%%%%%%%%%%%%%%%%%%%%%%%%%%%%%%%%%%%
Here we discuss the modifications of the short-distance nuclear-
reaction rates mainly governed by the strong interaction.  In CHAMP
BBN (CBBN), the corresponding dominant process for  the SBBN forward
process $a+b\to c+d$ may be $(a,C)+b\to (c,C)+d$ or $c+d+C$, assuming
$(b,C)$ does not  have sufficiently a large binding energy against
scattering of background photons, i.e., $E_{\text{bin}}/T\ll 40$. Here
$(c,C)$ has a larger binding energy than that of $(d,C)$. If the
following condition is satisfied,
\begin{eqnarray} 
&&{\it Q}_{\text{SBBN}}-E_{\text{bin},aC}>0,
\end{eqnarray}
the final state is given by $(a,C)+b\to c+d+C$.

On the other hand, even if the above condition is not satisfied, but
if the following condition is satisfied,
\begin{eqnarray} 
{\it Q}_{\text{CBBN}}={\it Q}_{\text{SBBN}}+E_{\text{bin},cC}-E_{\text{bin},aC}>0 ,
\end{eqnarray}
$(a,C)+b\to (c,C)+d$ is kinematically allowed, and  the CHAMP in the
final state will be trapped again.  However if the bound state $(c,C)$
does not have enough binding energy  against the destruction due
to thermal photons, the $(c,C)$ state will be destroyed soon after the
process, and the element $c$ and the  CHAMP will become free.

For the $Z_C=1$ case and the relevant nuclei,  because most of the
${\it Q}_{\text{SBBN}}$ values are sufficiently large, the case that
${\it Q}_{\text{SBBN}}>0$ but ${\it Q}_{\text{CBBN}}<0$ would be rare.  However,
in general, it might be possible. In such cases,  even though the SBBN
process does not have any threshold, the CBBN can have it. But  the
sign flip in the {\it Q} value occurs when the binding energy of a bound state with 
a CHAMP exceeds the nuclear binding energy of the process, which may mean 
that the bound CHAMP is not a spectator in the nuclear-reaction any more. 
In our following analysis, we do not consider this kind of special cases.
 
Next we simply assume that ${\it Q}_{\text{SBBN}}{\it Q}_{\text{CBBN}}>0$. Let us
consider the reverse processes  of $a+b\to c+d$ in CBBN, which has a
threshold characterized by ${\it Q}_{\text{CBBN}}$.  Then, the possible
dominant process would be the SBBN process  $c+d\to a+b$ if $(c,C$)
and $(d,C)$ are not stable against scattering off the background
photons. In addition, $(c,C)+d\to (a,C)+b$ can be also another
dominant process if $(d,C)$ is not stable in the thermal bath, for
simplicity  assuming $(a,C)$ has larger binding energy than $(b,C)$.
\footnote{$(c,C)+d\to a+b+C$ is also possible if it is kinematically
allowed.}  In these processes,  we may expect a Boltzmann suppression
factor in the reaction rate $e^{-{\it Q}_{\text{CBBN}}/T}$, not
$e^{-{\it Q}_{\text{SBBN}}/T}$ in a similar fashion in SBBN.

If the SBBN strong interaction $a+b\to c+d$ occurs at a shorter
time scale than the typical time scale of electro-magnetic (EM) interactions of the bound
states,  we may expect that such a  short-distance reaction rate
should not be deviated from the SBBN rate. For D, T, He, Li, Be etc,
this condition can be realized easily.

%%%%%%%%%%%%%%%%%%%%%%%%%%%%%%%%%%%%%%%%%%%%%%%%%%%%%%%%%%%%%%%%%%%%%%
\subsubsection{Flux}
%%%%%%%%%%%%%%%%%%%%%%%%%%%%%%%%%%%%%%%%%%%%%%%%%%%%%%%%%%%%%%%%%%%%%%
In general, the velocity $V_{\rm flux}$ which controls the flux might be
different from the velocity $V_{\rm reac}$ which controls the short
distance  nuclear reaction. The $\langle
\sigma_{\text{fusion}}V_{\rm flux} \rangle $ would be given by
\begin{eqnarray}
\langle \sigma_{\text{fusion}}V_{\rm flux} \rangle =
(\sigma_S\frac{V_{\rm flux}}
{V_{\rm reac}}+\sigma_P(V_{\rm flux}V_{\rm reac})....)
\label{eq:sigmaVf}
\end{eqnarray}
Here, we assume that for the 
short distance reactions, the coefficients, $\sigma_S$, $\sigma_P$...
in CBBN are the same as in SBBN
\footnote{If the  phase space  is modified by  the release of  a CHAMP
after the  reaction, the difference from  the SBBN case  would be also
small if the  Q-value is large.}. Using this approximation, 
we evaluate the flux.

First, we consider collisions between a bound state and a free light
element. Then, once we focus on the $2\to 2$ collision between the
bound and the free light element, the relative velocity $V_1$ may be
dominated  by the speed of the bound light element. If we assume that
the free light  element is distributed uniformly in the thermal bath, 
the flux is controlled  by $V_1$. On the other hand, in the case that the
radius of the  bound state is smaller than the impact parameter of
nuclear reactions [which is $O(1/m_{\pi})$],  the flux has to be
estimated by the relative velocity between the  bound state and the
free light element, which is controlled by the relative velocity. But
even in such cases, $V_{\rm flux}\sim V_{\rm reac}$ due to 
the following consideration. Taking $V_{\rm reac}=V_1$,  while the
free element goes through the target volume, the bound light element
rotates with the speed $V_1=\sqrt{2E_{\text{bin}}/m_X}$. Then the
number of rotations would be  $\sim V_1\Delta t/2\pi r_B\sim
O(V_1/V_2)$ where $\Delta t\sim 2r_B/V_2$ is the time for the free
light  element to go through the bound light element, $V_2$ is the
velocity of  the free light element, and $r_B$ is the radius of the
bound state.  \footnote{This discussion rely on an assumption that the
factorization of Coulomb factor and short-distance nuclear fusion is
valid. That is, we assumed that in the collision, the bound state  is
not destroyed before the collision. This would be valid if $r_B \sim
1/m_{\pi}$. If the bound state is unstable against incident nucleus,
the effective $V_{\rm flux}/V_{\rm reac}$ may become smaller than
unity.}  Then, for the nuclear reaction due to pion exchange, if
we take  $V_{\rm reac}=V_1$, the flux is the relative velocity $V_2$
times $O(V_1/V_2)$ which would be $\sim V_1$

Next, we consider collisions between a neutral bound state $a$ and a
neutral (or charged) bound state $b$. In this case, since the target is
not a freely propagating particle, the speed which controls  the flux
is not the  bound light element's $V_1\sim V_a+V_b$ but the relative
velocity $V_2$ between the bound states. $V_2$ is order of the thermal
velocity  of the bound state, which is smaller than $V_1$. Then
$V_2/V_1\sim O(0.1)$  at around $T$=1keV where neutral bound states
can be formed. However while the bound states collide with each
other, the bound element would rotate around a CHAMP $\sim (V_1/V_2)$
times.  Therefore even in this case, we could estimate $V_{\rm
flux}\sim V_1$.

These considerations imply that we can simply assume that the CHAMP in the
bound states is a spectator and $V_{\rm reac}\simeq V_{\rm
flux}$.  \footnote{Our consideration is based on our approximation
that the short-distance reaction is the same as that of the SBBN $2\to
2$ process between light elements. In the case that the De-Broglie 
wavelength of an incoming nucleus is longer than the Bohr radius of the 
bound state, we may have to solve quantum mechanical many-body problems including
 a bound CHAMP to obtain a more reliable result.}

%%%%%%%%%%%%%%%%%%%%%%%%%%%%%%%%%%%%%%%%%%%%%%%%%%%%%%%%%%%%%%%%%%%%%%
\subsubsection{Corrections for BBN nuclear reaction rates with no threshold}
%%%%%%%%%%%%%%%%%%%%%%%%%%%%%%%%%%%%%%%%%%%%%%%%%%%%%%%%%%%%%%%%%%%%%%
Here we consider the nuclear reactions containing light elements
captured by CHAMPs. In this case, as we mentioned before, the crucial
differences from SBBN are in the Coulomb factor and the Boltzmann
suppression.  Since the radius of bound state is very small
$O(1/m_{\pi})$,  a simple replacement $Z_X\to Z_{(X,C)}=Z_X-Z_C$ in the
Coulomb factor would be a good approximation. However the short
distance part is also changed  because the light element captured by a
CHAMP has the kinetic energy  $E_{\text{bin}}$ not $O(T)$. As we
mentioned before,  we assume that the  short-distance cross section
$\sigma_{\text{fusion}}^C$  takes the same functional form of the CM
energy as those of SBBN, $\sigma_{\text{fusion}}$.  Thus the CM energy
of the short-distance nuclear reaction may be
$O(Max(E_{\text{bin}},E_0))$. We introduce these two changes in the
estimation of nuclear reaction rates. That is,
\begin{eqnarray}
\sigma_{\text{fusion}}^CV&=&(\sigma_S^C+\sigma_P^C V^2 +....)F_{(aC)b}(\beta)
\nonumber\\
&\simeq&\sigma_0 v(V)\frac{2\pi Z_{(aC)}Z_b\alpha}{\beta} 
e^{-\frac{2\pi Z_{(aC)}Z_b\alpha}{\beta}}\nonumber\\
\end{eqnarray}
where $\beta$ is the relative velocity between the bound state and the
incident thermal light element,  and $V$ is the relative velocity
between the bound light element and  the incident thermal light
element with $E=E_0$.  $\beta$ controls the amount of penetration in the 
Coulomb potential. $V$ appears in the flux and the short  distance
cross section.

Since the short distance cross section would be governed by  the
kinetic energy of the bound light element which does not depend on the
condition of the thermal bath much, the thermal average should be taken
only for the Coulomb part which implies the evaluation of
the wave function  for an incident thermal light element at the position
of a bound state.  Then the thermal average may be taken for the
thermal light elements and  the thermal bound state  because the
incident thermal light element approaches inside the  Coulomb field of
the bound state, not that of bound light elements.  We assume that the
short distance reaction is faster than the  EM interaction of the bound
state.  The thermal-averaged cross section is calculated as follows.
\begin{eqnarray}
&& \langle \sigma_{\text{fusion}}^C V \rangle \nonumber\\
&=&
\frac{g^2}{(2\pi)^6n_bn_{(aC)}}
\int d^3p_{(aC)} d^3p_b \sigma_0 v(V)
\frac{2\pi Z_b Z_{(aC)}\alpha}{\beta}
\nonumber\\
&&~~~~~\times
e^{-\frac{2\pi Z_b Z_{(aC)}\alpha}{\beta}}
e^{-\frac{(E_{(aC)}+E_b)}{T}}, \nonumber\\
&=&
\frac{g}{(2\pi)^3n_r}\int d^3p_r \sigma_0 v(V)\frac{2\pi Z_bZ_{(aC)}\alpha
\mu_{(aC)b}}{p_r} \nonumber\\
&&~~~~~\times e^{-\frac{2\pi Z_b Z_{(aC)}\alpha\mu_{(aC)b}}{p_r}}
e^{-\frac{E_r}{T}}, \nonumber\\
&=&\frac{8\pi gT\mu_{(aC)b}}{(2\pi)^3n_r}\int dy S(y T)_{\text{New}}
e^{-(y+\sqrt{\frac{y_G}{y}})},
\end{eqnarray}
where $S(y_XT)_{\text{New}}=\sigma_0 v(V) \pi Z_b Z_{(aC)}\alpha
\mu_{(aC)b}$,  $y_G=2\pi^2Z_b^2 Z_{(aC)}^2\mu_{(aC)b}\alpha^2/T$,
$y_0=(y_G/4)^{1/3}$ and  $\mu_{(aC)b}=m_{(aC)}m_b/(m_{(aC)}+m_b)\simeq
m_b$.  Notice that we are assuming that the short-distance nuclear
cross sections have the same functional forms  of the CM energy as
those of SBBN.

Here in the case of $Z_{(aC)}\neq 0$, we relate the new S-factor above to the
SBBN {\it S} factor which could be measured by experiments,
\begin{eqnarray}
S(y_X T)_{\text{New}}=S(y_X T)\frac{Z_{(aC)}}{Z_a}\frac{\mu_{(aC)b}}{\mu_{ab}}.
\end{eqnarray} 
Then we find
\begin{eqnarray}
&& \langle \sigma_{\text{fusion}}^C V \rangle (T)
=\nonumber\\
&&~~~~~\sqrt{\frac{32}{4^{1/3}}\frac{\tilde{E}_G^{1/3}}{3\mu_{(aC)b}}}
\frac{S(y_XT)_{\text{New}}}{T^{2/3}}
e^{-\frac{3}{4^{1/3}}(\frac{\tilde{E}_G}{T})^{1/3}},
\end{eqnarray}
where $y_X\simeq (((\mu_{ab}/m_a)E_{\text{bin}})+\tilde{E}_0)/T
\sim (E_{\text{bin}}+\tilde{E}_0)/T )$,
 $\tilde{E}_G=2\pi^2Z_b^2 Z_{(aC)}^2\mu_{(aC)b}\alpha^2$ and 
$\tilde{E}_0=Ty_0$.

For nuclear reaction rates with neutrons like $^7$Be(n,p)$^7$Li, since
there is no Coulomb suppression or Boltzmann suppression if  there is
no threshold in the process [$(aC)+n\to (cC)+d$],  we replace CM
energy by $E_{\text{CM}}=(\mu_{ab}/\mu_{(aC)b})E_{\text{bin}}+3T/2
\sim E_{\text{bin}}+3T/2$ in the cross sections because of the change
of the kinematics of the bound light elements. In addition if the
bound state is neutral ($Z_{(aC)}= 0$), the Coulomb factor may
disappear if the bound state is not destroyed before the
collision. Then the treatments may  be similar to the neutron
case above.  Such neutral bound states will be formed in  case of  $Z_C=1$
(proton, D, and T).

In the above discussions, we have taken the approximations that the light
element is pointlike and does not have internal structure, 
and the selection rules in the nuclear reactions are
not  changed by the trapped  CHAMP.~\footnote{We have also assumed the
hierarchy between the SBBN strong reactions and the EM interactions of
the bound states. The Bohr radius of CHAMP-light element system and
the typical pion-exchange radius would be the same order of magnitude
in our case, but we can still expect a hierarchy in the coupling
strengths between EM and strong interaction, which may still allow us
to factorize the short distance nuclear reaction from the effects
caused by binding  a CHAMP.But if the incoming nucleus is very slow, this 
factorization may break down due to the long range nature of the EM force}

%%%%%%%%%%%%%%%%%%%%%%%%%%%%%%%%%%%%%%%%%%%%%%%%%%%%%%%%%%%%%%%%%%%%%%
\subsubsection{Corrections for BBN nuclear reaction rates with threshold}
%%%%%%%%%%%%%%%%%%%%%%%%%%%%%%%%%%%%%%%%%%%%%%%%%%%%%%%%%%%%%%%%%%%%%%
Let us consider a case that the SBBN reverse process $c+d\to a+b$ has
a threshold and the SBBN cross section of $a+b\to c+d$ can be measured
by collider experiments.

First, assuming the condition $E_{\text{bin}(aC):1S}<
Q_{\text{SBBN};ab,cd}<E_{\text{bin}(aC):1S}+E_{\text{bin}:1E}$ is satisfied 
where $E_{\text{bin}(aC):1S}$, $E_{\text{bin}:1E}$ are the binding energies
of  the 1S state of the $(a,C)$ system and of the first excited level of
the $(c,C)$ system,  we can estimate the cross section of $(c,C)+d
\to(a,C)+b$ by using the information of the SBBN forward process
$a+b\to c+d$. Under the above conditions, we may use the
detailed balance relation on $(a,C)+b\to (c,C)+d$  in a similar
fashion to the previous discussion.  The thermal-averaged cross
section of $(c,C)+d\to (a,C)+b$ may be written as follows.  Applying
the detailed balance relation and the modifications for  the forward
process which was previously discussed,
\begin{eqnarray}
&& \langle \sigma_{\text{fusion},(cC)d}^CV \rangle \nonumber\\
&&\simeq\frac{g_{(aC)}g_b}{g_{(cC)}g_d}(\frac{m_b}{m_d})^{3/2}
\langle \sigma_{\text{fusion},(aC)b}V \rangle
e^{-\frac{Q_{\text{CBBN}ab,cd}}{T}} 
\end{eqnarray}
where ${\it Q}_{\text{CBBN}ab,cd}={\it Q}_{(aC)b,(cC)d}$. If
${\it Q}_{\text{CBBN}ab,cd}$ is small, the Boltzmann suppression might
disappear even though SBBN has a large Boltzmann suppression.

For ${\it Q}_{\text{CBBN}:ab,cd}>E_{\text{bin}(cC):2E}$ where
$E_{\text{bin}(cC):2E}$ is the binding energy of the second excited
level of the bound state, we would not be able to simply apply the
detailed balance relation for the forward process.  But in any case,
since the crucial point for the processes with a threshold is the
Boltzmann suppression which comes from the requirement that the kinetic
energy of the incident particles overcomes the threshold,  if the
{\it Q} value is smaller than that of SBBN, we may expect the milder
Boltzmann suppression in the process, compared to that of
SBBN.~\footnote{Here we naively assumed  ${\it Q}_{\text{CBBN}:ab,cd}>E_0$
where $E_0$ is the Gamow peak energy of the reverse  processes. If
this condition is not satisfied, we may need more careful treatments.}

In the $Z_C=1$ case, at a relevant time when capture become efficient,
the Boltzmann suppression is huge if the {\it Q} value is O(MeV), and then
most of the BBN processes are completely decoupled.  Hence we
ignored the change of nuclear reaction rates for the SBBN reverse
processes if ${\it Q}_{\text{CBBN}}$ is $\sim \cal{O}$(1)MeV, which is a 
reasonable assumption.

Next, we consider the reverse process in CBBN, which corresponds to
SBBN $a+b\to c+\gamma$, i.e.,  $(c,C)+\gamma\to (a,C)+b$ assuming that
the binding energy of $(a,C)$ is smaller than that of $(b,C)$.  It is
well-known that the reaction rate of this forward process is small.
Notice that the incident photon with  the threshold energy of the
process does not have Coulomb suppression.  Thus the main origin of
suppression is the low abundance of the higher energy components of
thermal photons.
\begin{eqnarray}
&& \langle \sigma_{\text{fusion},c\gamma}^CV \rangle 
\nonumber\\&=&
\frac{8\pi g_{\gamma}}{(2\pi)^3n_r}
\int dE_{r,(cC)\gamma} E_{r,(cC)\gamma}^2
\sigma_{0,c\gamma}v(V) 
e^{-\frac{E_{r,(cC)\gamma}}{T}}, \nonumber\\
&\simeq&
\frac{8\pi g_{\gamma}}{(2\pi)^3n_{\gamma}}
\int_{Q_{\text{CBBN}}}^{\infty} dp_{\gamma} p_{\gamma}^2 
\sigma_{0,c\gamma}v(V)e^{-\frac{p_{\gamma}}{T}},
\nonumber\\
&\simeq&\frac{1}{n_{\gamma}}
(\frac{\mu_{(aC)b}T}{2\pi})^{3/2}O(
\langle \sigma_{\text{fusion},(aC)b}v \rangle )e^{-\frac{Q_{\text{CBBN}}}{T}},
\end{eqnarray}
where $E_{r,(cC)\gamma}=p_{\gamma}$.  Although the {\it Q} value for the
process $^3$He($\alpha$,$\gamma$)$^7$Be might be smaller than
1MeV, the process is negligible at the capture time of  $^7$Be and
this reverse process does not seem to provide a significant change  from
SBBN.  The change on the  threshold energy of these photodissociation
processes might be important when we consider the late-decay effects
that  the injected high energy EM energy is thermalized and produce
a huge number of soft photons, which may destroy primordial light
elements.

%%%%%%%%%%%%%%%%%%%%%%%%%%%%%%%%%%%%%%%%%%%%%%%%%%%%%%%%%%%%%%%%%%%%%%
\section{BBN with long-lived CHAMPs}
%%%%%%%%%%%%%%%%%%%%%%%%%%%%%%%%%%%%%%%%%%%%%%%%%%%%%%%%%%%%%%%%%%%%%%
Recently WMAP has reported the updated values of cosmological
parameters under the standard $\Lambda$CDM models. We can now check 
the internal consistency of  SBBN  in the light of WMAP3. 
It has been pointed out that the predicted $^7$Li abundance seems 
too high to agree with observed abundances. Also for $^6$Li, 
we have to expect an additional production after the BBN epoch,
like cosmic-ray nucleosynthesis. These tensions or discrepancies may
be tantalizing clues to find new physics. Under these circumstances,
it is interesting to study the effects of new physics.

In previous sections, we considered the possible changes of nuclear
reaction  rates due to long-lived CHAMPs. Here we consider the
application for  the BBN in  case of $Z_C=1$.

%%%%%%%%%%%%%%%%%%%%%%%%%%%%%%%%%%%%%%%%%%%%%%%%%%%%%%%%%%%%%%%%%%%%%%
\subsection{Charged Massive Particle BBN (CBBN)}
%%%%%%%%%%%%%%%%%%%%%%%%%%%%%%%%%%%%%%%%%%%%%%%%%%%%%%%%%%%%%%%%%%%%%%
We consider the thermal freeze-out of light element abundances in CBBN
and here we simply ignore the effects of possible high-energy
injections due to the late decay of CHAMPs,  which may provide the
initial condition to consider such late decay phenomenon if the decay
occurs long enough after the decoupling of  the BBN processes.  We will
later discuss  the case where the decays occur before the freeze-out. 
In our estimation, we also assume the instantaneous captures
for each light elements at $T_c=E_{\text{bin}}/40$.  \footnote{  As we
mentioned before, if the number density of CHAMPs is low
$n_{\text{CHAMP}}/n_{\gamma}\ll 10^{-11}$,  the recombination rate
might  not  be sufficiently large compared with  the expansion rate of
the Universe, and we may expect poor captures of CHAMPs.  Then the
most of CHAMPs and light elements will be left as freely-propagating
ionized particles. Because CHAMPs are supposed to decay soon, in this
case we can apply the known results in decaying particle scenarios in
literature. }

In SBBN, abundances of all light elements are
completely frozen until $T\sim$ 30keV. Since  $T_c$ is 24keV for
$^7$Li and 38keV for $^7$Be, which is almost the end  of  SBBN, 
the formations of bound states may change their abundances. For
elements lighter than $^6$Li, since the efficient captures  occur only
at below 10 keV,  we found that the change of nuclear reactions can
not recover the processes  at such a low temperature.  This conclusion
will hold if the difference from our estimation  of $\langle
\sigma_{\text{fusion}}V \rangle $ is not large. Also in most of
the reverse processes,  the Boltzmann suppressions are huge at that time,
even though we use the new  {\it Q} value ${\it Q}_{\text{CBBN}}$. They do not
provide any significant change from SBBN.

Under these circumstances, if the CHAMPs decay before the captures of
$Z_X=1$ nuclei,  we may expect that the sizable change due to the
captures occur  in elements heavier than $^7$Li.  On the other hand,
once the capture of proton, D and T starts,  since the bound states
are neutral and have no Coulomb suppressions in  the nuclear
reactions, the BBN processes may not freeze out. In the next
subsection, first of all, we consider the case that CHAMPs decay
before the captures of $Z_{X}=1$ elements such as proton, D, or T,
which  start at below $T\lesssim$~1--2~keV ($t\gtrsim 10^6$sec).
Later we consider the possible effects due to their captures.

Since the abundances of the light elements differ by orders of magnitude, 
often we can identify the relevant processes and neglect the others.  For
example, when we are considering a process $a(b,c)d$, if $n_a$ is much
smaller than the others ($n_b$, $n_c$ and $n_d$), this process is
negligible for the evolutions of $n_b$, $n_c$ and $n_d$, but important
only for $n_a$. Therefore elements heavier than $^7$Be do not significantly 
affect lighter elements abundances.

%%%%%%%%%%%%%%%%%%%%%%%%%%%%%%%%%%%%%%%%%%%%%%%%%%%%%%%%%%%%%%%%%%%%%%
\subsubsection{CBBN with $Z_X>1$}
\begin{figure}[tbp]
\postscript{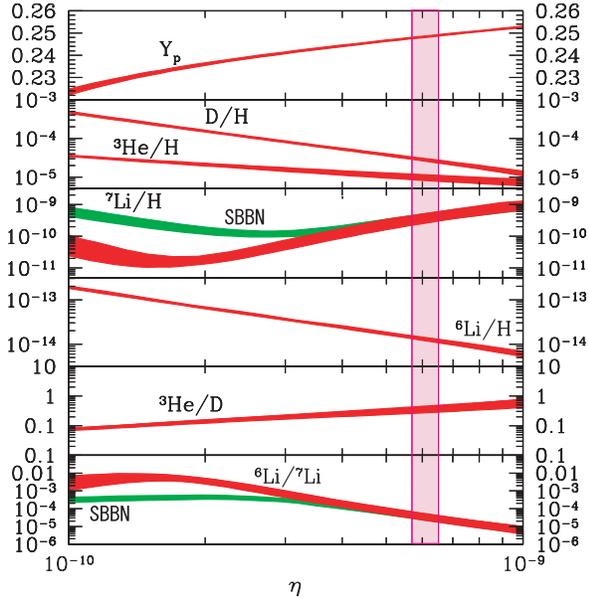}{0.95}
\caption{Theoretical predictions of $Y_p$, D/H, $^3$He/H, $^7$Li/H
$^6$Li/H, $^3$He/D and $^6$Li/$^7$Li   as a function of $\eta$ in
standard  BBN (green) and CHAMP BBN in case A (red). Here we have
assumed the instantaneous capture of CHAMPs and
$n_{\text{C}}/n_{\gamma}=3.0\times 10^{-11}$. 
\label{fig:yield_ebin40_cl} }
\end{figure}

\begin{figure}[tbp]
\postscript{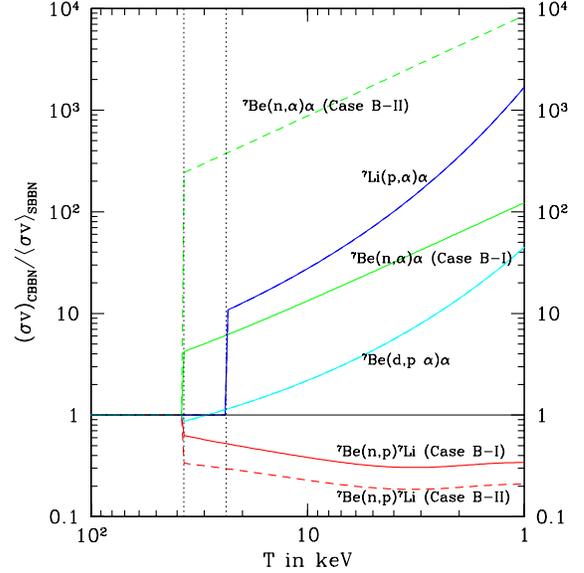}{0.95}
\caption{Ratio of  nuclear-reaction rates
of SBBN and CBBN  in Case B-I and B-II  as a function of the cosmic
temperature for the relevant processes. Here we assumed  the
instantaneous capture of CHAMPs by the  nuclei.  Case B-I means
that $E_{\text{CM}}=(\mu_{ab} /\mu_{(aC)b})E_{\text{bin}}+E_0$ in a
process $(a,C)+b\to (c,C)+d$ where we take $E_0$ to be  the Gamow's
peak energy for collisions between two charged elements, and to be
$3T/2$ for collisions between a nucleus and a neutron. The Case B-II
means that we take $E_{\text{CM}}=E_{\text{bin}} +E_0$ as the CM
energy of processes and 10 times larger value of the p-wave part of
the cross section of $^7$Be(n,$\alpha$)$^4$He than that in the
standard BBN code~\cite{Wagoner:1969,Serpico:2004gx}.
\label{fig:rates} }
\end{figure}

%%%%%%%%%%%%%%%%%%%%%%%%%%%%%%%%%%%%%%%%%%%%%%%%%%%%%%%%%%%%%%%%%%%%%%
Here we consider the CBBN with captures of $Z_X>1$ nuclei (with
$Z_C=1$).  This case will be realized if the CHAMP lifetime is shorter
than $\sim 10^6$sec.  In Fig.~\ref{fig:yield_ebin40_cl}, we show a plot
of the light element abundances as a function of $\eta$, including the
corrections only in processes among charged light elements
(Case~A). We can find that the $^7$Li abundance could decrease much
from the SBBN value for $\eta$.  The decrease is induced by the
enhancement of $^7$Li(p,$\alpha$)$^4$He reaction rate due to the
capture of $^7$Li by CHAMPs. As we can see in Fig.~\ref{fig:rates},
the CBBN reaction rate of $^7$Li(p,$\alpha$)$^4$He slowly decreases as
a function of the energy, compared to that of SBBN at the temperature
where the Coulomb suppression becomes important, which results
in later-time decoupling of the process  than in SBBN.

We also added processes $^7$Be(n, p)$^7$Li and
$^7$Be(n,$\alpha$)$^4$He, which are associated with neutron capture
(Case~B). In these type of processes, the important change from  SBBN
is the kinetic energy to be used in the nuclear reaction. In SBBN, the
typical energy is $\sim 3T/2$. However in CBBN, the energy could be
$O(E_{\text{bin}})$. If the s-wave partial wave mode dominates the
process, then the difference might be small. However, if higher
partial modes such as p-wave dominate, we expect significant
enhancements of the processes.

In fact, we found that the change in $^7$Be+$^7$Li by the modification
of the process  $^7$Be(n,p)$^7$Li is negligible. On the other hand,
since $^7$Be(n,$\alpha$)$^4$He is a p-wave dominant
process~\cite{Wagoner:1969}, the modification of this process should be
important to predict the primordial abundance  of $^7$Be+$^7$Li in
CBBN.  Unfortunately, we currently only have poor experimental data sets for
$^7$Be(n,$\alpha$)$^4$He. However, since there is experimental data
for the reverse process $^4$He($\alpha$,n)$^7$Be~\cite{2alpha},
we might be able to theoretically infer the cross section of the
forward process of  $^7$Be(n,$\alpha$)$^4$He approximately by using
detailed balance relations. For the moment, however, the experimental
data do not have sufficient resolutions in the relevant energy region
because of the significant Coulomb suppression and the threshold
suppression, to correctly calculate the forward rate.  Therefore,
according to Serpico et al.~\cite{Serpico:2004gx},  as a conservative
error we also take a factor of 10  on the process in this paper,
which does not change the SBBN predictions at all and is still
consistent with available experimental data of the  reverse
rate~\cite{2alpha}.  

In Fig.5, we plot the theoretical
prediction of $^7$Li/H ({\it upper panel}) and $^6$Li/$^7$Li ({\it
lower  panel}) as a function of $\eta$. The SBBN predictions are marked
by the green bands. The red (blue) band is for Case B-I (Case B-II) in
CBBN. Here we assumed $n_{\text{C}}/n_{\gamma}=3.0\times 10^{-11}$ and
the instantaneous capture of CHAMPs. Case B-I means that
$E_{\text{CM}}=(\mu_{ab} /\mu_{(aC)b})E_{\text{bin}}+E_0$ in a process
$(a,C)+b\to (c,C)+d$ where we take $E_0$ to be  the Gamow peak
energy for collisions between two charged elements, and to be $3T/2$
for collisions between a nucleus and a neutron. Case B-II means
that we take $E_{\text{CM}}=E_{\text{bin}} +E_0$ as the CM energy of
processes and a 10 times larger value of the p-wave part of the cross
section of $^7$Be(n,$\alpha$)$^4$He than that in the standard BBN
code~\cite{Wagoner:1969,Serpico:2004gx}. In
Fig.5, it is showed that the modification
by a factor of 10 on the p-wave partial cross section of
$^7$Be(n,$\alpha$)$^4$He does not change the  SBBN prediction (Case
B-I) but must be important in CBBN (Case B-II).

We have also checked the reverse process of
$^{10}$B(p,$\alpha$)$^7$Be.  The threshold in this process can become
smaller, which may induce milder Boltzmann suppression than that of
SBBN.  However, we found that this rate is simultaneously suppressed
strongly by the Coulomb factor, and therefore this effect is
irrelevant.

Finally we warn the readers again that  our results
rely on the assumption that the short-distance nuclear-reaction rates
have the same functional form of the  CM energy as those of SBBN. In
addition, we assume that by relevant elements, the energy
to excite nucleons into higher levels and the  binding energy by a
CHAMP are of the same order of magnitude. To obtain a quantitative
conclusion, further efforts to estimate the errors in the
short-distance nuclear reaction rates must be important.  For example,
in $^7$Be(n,p)$^7$Li, the change of the nuclear reaction rate  can
directly affect on the final abundance of $^7$Li(=$^7$Li+$^7$Be).
However notice that well before the elements lighter than Li are
captured by CHAMPs,  the SBBN processes are completely decoupled. Even
though the errors induce larger reaction rate, if it were within an
order of magnitude level, nuclear reaction would not overcome  the
expansion rate again, and our conclusion would not be changed, because
the Coulomb suppression is significant, and the neutron abundance is
very small.

%%%%%%%%%%%%%%%%%%%%%%%%%%%%%%%%%%%%%%%%%%%%%%%%%%%%%%%%%%%%%%%%%%%%%%
%%%%%%%%%%%%%%%%%%%%%%%%%%%%%%%%%%%%%%%%%%%%%%%%%%%%%%%%%%%%%%%%%%%%%%

%%%%%%%%%%%%%%%%%%%%%%%%%%%%%%%%%%%%%%%%%%%%%%%%%%%%%%%%%%%%%%%%%%%%%%
%%%%%%%%%%%%%%%%%%%%%%%%%%%%%%%%%%%%%%%%%%%%%%%%%%%%%%%%%%%%%%%%%%%%%%

%%%%%%%%%%%%%%%%%%%%%%%%%%%%%%%%%%%%%%%%%%%%%%%%%%%%%%%%%%%%%%%%%%%%%%
\begin{figure}[tbp]
\postscript{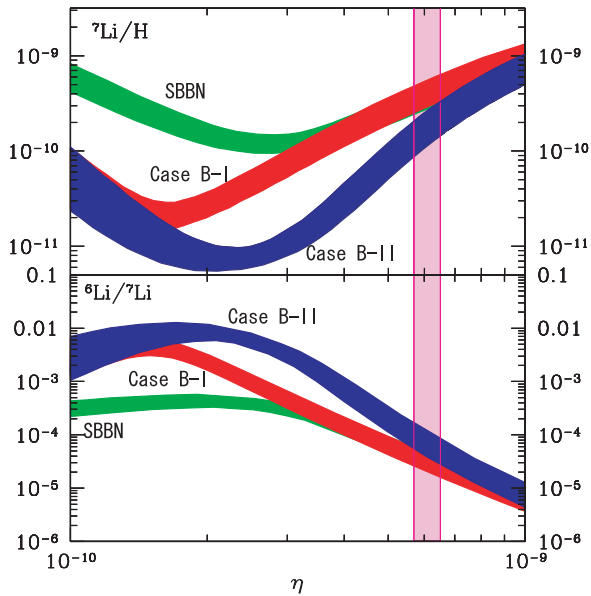}{0.95}
\caption{Theoretical prediction of $^7$Li/H ({\it upper panel}) and
$^6$Li/$^7$Li ({\it lower  panel}) as a function of the
baryon-to-photon ratio. The SBBN predictions are marked by the green
bands. The red (blue) band is for Case B-I (Case B-II) in CBBN. Here
we assumed $n_{\text{C}}/n_{\gamma}=3.0\times 10^{-11}$  and the
instantaneous capture of CHAMPs. The definition of  Case B-I and Case
B-II are same as those in Fig.~\ref{fig:rates}.
\label{fig:ield_ebin40_n_large} }
\end{figure}
%%%%%%%%%%%%%%%%%%%%%%%%%%%%%%%%%%%%%%%%%%%%%%%%%%%%%%%%%%%%%%%%%%%%%%

%%%%%%%%%%%%%%%%%%%%%%%%%%%%%%%%%%%%%%%%%%%%%%%%%%%%%%%%%%%%%%%%%%%%%%
\subsubsection{CBBN with $Z_X=1$ }
%%%%%%%%%%%%%%%%%%%%%%%%%%%%%%%%%%%%%%%%%%%%%%%%%%%%%%%%%%%%%%%%%%%%%%
Next we discuss the possible effects due to captures of  $Z_X=1$
nuclei. Since the bound states are neutral,  the nuclear reactions in
TABLE II may not have Coulomb suppression and might be significantly
changed from those of SBBN.  ~\footnote{Notice that here we simply
assumed that the bound state is not  significantly disturbed before
the nuclear fusion reactions.}

We consider the case that $Z_C=1$ nuclei are captured instantaneously
at temperatures below each  $T_c~(\sim$~\cal{O}(1)~keV).  The  captures of 
T only provide a significant change in  T itself  and $^7$Li even if we assume 
the instantaneous captures because of their poor  abundances.  The
captures of D result in large enhancements  for the processes listed in
TABLE II. In particular, since T(d,n)$^4$He,  $^3$He(d,p)$^4$He,
$^7$Li(d,n$\alpha$)$^4$He and $^7$Be(d,p$\alpha$)$^4$He have large
cross sections, the reaction rates may be able to  become larger than
the expansion rate again at a later time.  Their decoupling does  not
occur   soon  because  of the  absence of the Coulomb suppressions.
If the  captured D abundance is larger than $^3$He, D and  $^3$He
mainly  burn into $^4$He through $^3$He(d,p)$^4$He. Then the abundance
of $^3$He can decrease, and the  abundance of D becomes close to
$n_D-n_{^3He}$. On the other hand, however,  D(d,p)T  and D(d,n)$^3$He
do not change the abundance of D so much.  In addition,  T(d,n)$^4$He,
$^7$Li(d,n$\alpha$)$^4$He, and $^7$Be(d,p$\alpha$)$^4$He do not a 
change the abundance of D either, but might decrease the abundances of T,
$^7$Li and $^7$Be because  of the small abundances compared with  that
of  D.

Although the process D($\alpha$,$\gamma$)$^6$Li has very small
reaction rate in SBBN since the  abundances of the incident particles
(D and $^{4}$He) are sufficiently large,  this process can produce
large amount of  $^6$Li.~\footnote{Recently 
Pospelov pointed out that the cross section of D($\alpha$,$\gamma$)$^6$Li 
might be significantly enhanced by considering the virtual photon absorption 
due to a bound CHAMP~\cite{Pospelov:2006sc}, which 
might significantly overproduce $^6$Li. Since his paper appeared after the 
completion of this work, we have not included this effect in this paper.}   
The capture of protons can reduce the  $^6$Li and
$^7$Li abundances through $^7$Li(p,$\alpha$)$^4$He and
$^6$Li(p,$\alpha$)$^3$He.  On the other hand, there are no significant
changes on D, T and $^7$Be abundances because  the associated
processes are radiative ones,  which are relatively suppressed.

\begin{figure}[tbp]
    \postscript{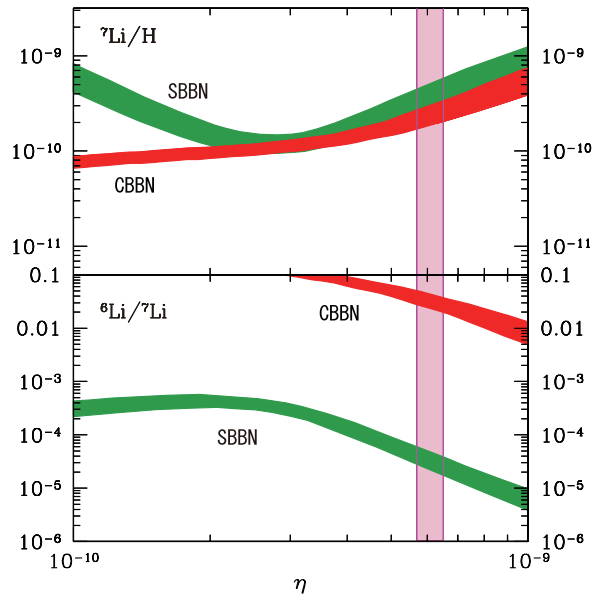}{0.95}
\caption{Theoretical predictions of 
$^7$Li/H and $^6$Li/$^7$Li as a function of $\eta$ in SBBN and CBBN
with their theoretical errors at 95 $\%$ C.L. Here we took the
fractions of the captured proton, D, and T to be  $10^{-5}$,
$10^{-1}$, and $10^{-2}$, respectively. The  WMAP value of $\eta$ at
95 $\%$ C.L. is also indicated as a vertical band.We can find that the
primordial values of $^6$Li and $^7$Li in the CBBN may be in the range 
of the observed abundances, which may simultanously solve current $^6$Li 
and $^7$Li problems pointed out in SBBN.
\label{fig:cap_D_a_g_Li6_m5_m1_m2_large}}
\end{figure}

In the relevant epoch for the captures of $p$, D and  T ($T \lesssim
\cal{O}$(1)~keV), the condition  to overcome the expansion rate is
that the reaction rates are larger than that of
$10^4\text{cm}^3/\text{sec}$ multiplied by the captured number density
of $n_p$.  Then in processes T(d,n)$^4$He, $^7$Li(d,n$\alpha$)$^4$He
and $^7$Be(d,p$\alpha$)$^4$He, even if the decrease of reaction rates
were within a factor of $O(10)$ due to some ambiguities such as
capture rate of D, we could still expect the decrease on $^7$Li and
$^7$Be abundances.  As we showed before, since the changes on light
element abundances  by the captures of $Z_X>1$ nuclei might be small,
the initial condition of light element abundances for such a later-time
CBBN by captured $Z_X=1$ nuclei might be the same as those of SBBN.
However notice that the above conclusions rely on  the number density of
the captured $Z_X=1$ nuclei very much. If the number density of
CHAMP is not large, the captures weaken, and  the changes become
milder.  For example, taking possible capture fractions, 
$O(10^{-5})$, $O(0.1)$, and $O(10^{-2})$ for proton, D, and T,
respectively, we show the results in
Fig.~\ref{fig:cap_D_a_g_Li6_m5_m1_m2_large}.  In such cases, the
nuclear-reaction rates for $^7$Li, $^6$Li and $^7$Be become more rapid
than the expansion of the universe, and we expect that $^7$Li and
$^7$Be decrease without  changing  D, $^3$He and $^4$He
abundances. The $^7$Li abundance is determined  by the competition
between two processes, $^7$Li(p,$\alpha$)$^4$He for proton capture and
T($\alpha$,$\gamma$)$^7$Li for T capture.  The $^6$Li is controlled by
the production reaction D($\alpha$,$\gamma$)$^6$Li, and the
destruction reaction $^6$Li(p,$\alpha$)$^3$He. In the case of
Fig.~\ref{fig:cap_D_a_g_Li6_m5_m1_m2_large}, a sizable amount of
$^6$Li is produced, and the predicted primordial value of
$^6$Li/$^7$Li approximately agrees with the  observational data
without assuming any chemical evolution scenarios.

%%%%%%%%%%%%%%%%%%%%%%%%%%%%%%%%%%%%%%%%%%%%%%%%%%%%%%%%%%%%%%%%%%%%%%
%%%%%%%%%%%%%%%%%%%%%%%%%%%%%%%%%%%%%%%%%%%%%%%%%%%%%%%%%%%%%%%%%%%%%%

On the other hand, notice that some ambiguities might still
exist in the nuclear-reaction rates. For $Z_X=1$ nuclei, because the
bound state with $Z_C=1$ CHAMP has a larger Bohr radius than
those of  $Z_X>2$ nuclei,  the electromagnetic disturbance on the
bound state before the nuclear fusion reactions occur would have to be
more carefully considered. If the bound state is electromagnetically
destroyed by an incident heavier nucleus, the factorization of Coulomb
part and short-distance nuclear reaction part does not work well, and
the nuclear-reaction rate may be changed from the value of our
calculations.~\footnote{For the collision of neutral bound state with
$Z_X=1$ nucleus, the formation of molecule may be important to
evaluate the nuclear-reaction rate.}  Also if a large amount of CHAMPs
survive until such late times ($>10^6$sec),  we would have to
simultaneously consider both effects due to the captures by the
$Z_X=1$ nuclei, and subsequently the EM energy injections by the
decaying CHAMPs at a later time.

%%%%%%%%%%%%%%%%%%%%%%%%%%%%%%%%%%%%%%%%%%%%%%%%%%%%%%%%%%%%%%%%%%%%%%
\begin{table}
\begin{tabular}{|rc|rc|rc|} \hline
process (p,C) & & The reaction rate (cm$^3$/sec/mol) & \\ \hline
$^7$Be((p,C),$\gamma$)($^8$B,C) & & $(4-24)\times 10^2$ & \\
$^7$Li((p,C),$\alpha$)($^4$He+C) & & $3 \times 10^6$ & \\
$^6$Li((p,C),$\alpha$)($^3$He+C) & & $1\times 10^8$ & \\
$^6$Li((p,C),$\gamma$)($^7$Be+C) & & $3\times 10^3$ & \\
T((p,C),$\gamma$)($^4$He+C) & & $2\times 10^2$ & \\
D((p,C),$\gamma$)($^3$He+C) & & $40$ & \\ \hline
process(D,C)  & & The reaction rate (cm$^3$/sec/mol) & \\ \hline
(D,C)(p,$\gamma$)($^3$He+C) & & $40$ & \\ 
(D,C)($\alpha$,$\gamma$)($^6$Li+C) & & 0.6 & \\ 
(D,C)(d,n)($^3$He+C) & & $7 \times 10^6$ & \\ 
(D,C)(d,p)(T+C) & & $4\times 10^6$ & \\ 
T((d,C),n)($^4$He+C) & & $1\times 10^9$ & \\ 
$^3$He((d,C),p)($^4$He+C) & & $4\times 10^{8}$ & \\ 
$^7$Li((d,C),n$\alpha$)($^4$He+C) & & $1\times 10^8$ & \\ 
$^7$Be((d,C),p$\alpha$)($^4$He+C) & & $3\times 10^8$ & \\ \hline 
processes(T,C)  & & The reaction rate (cm$^3$/sec/mol) & \\ \hline
(T,C)(p,$\gamma$)($^3$He+C) & & $2\times 10^2$ & \\ 
(T,C)(d,n)($^4$He+C) & & $1 \times 10^9$ & \\ 
(T,C)($\alpha$,$\gamma$)($^7$Li+C) & & $2\times 10^3$ & \\ \hline
\end{tabular}
\caption{Table of typical values for nuclear-reaction rates with a captured 
proton, D, and T. Here we ignored nuclear-reaction rates related to
the neutron because they are negligibly smaller for $T < T_{c}$. Here we also
ignored processes which include elements heavier than $^7$Be in the
initial state.}
\end{table}
%%%%%%%%%%%%%%%%%%%%%%%%%%%%%%%%%%%%%%%%%%%%%%%%%%%%%%%%%%%%%%%%%%%%%%

%%%%%%%%%%%%%%%%%%%%%%%%%%%%%%%%%%%%%%%%%%%%%%%%%%%%%%%%%%%%%%%%%%%%%%
\subsection{Late decays of long-lived CHAMP}
%%%%%%%%%%%%%%%%%%%%%%%%%%%%%%%%%%%%%%%%%%%%%%%%%%%%%%%%%%%%%%%%%%%%%%
We have discussed the change of light element abundances before the
decay  of CHAMPs. On the other hand, the decaying CHAMPs might induce
additional changes of primordial light element abundances,  which have
been studied by several groups.  The effects highly depend on the
decay products, i.e., electromagnetic or hadronic
cascades~\cite{Kawasaki:2004yh,Kawasaki:2004qu,HadOld,EMold}.

At an earlier epoch before $t = 10^4$sec, only the hadronic energy
injection is important, and there is almost no constraint from EM
energy injections.  Therefore, at such an epoch, even though the
injected energy is not small, if the branching ratio into hadronic
cascade is sufficiently suppressed, there are no significant effects
on the primordial light element abundances. Such a case is well known
if CHAMPs decay into leptons with a branching ratio into
hadrons of the order of $\cal{O}$($10^{-3}$)--$\cal{O}$($10^{-6}$).
For late decays after $10^4$ sec, the amount of energy release may be
highly constrained by the EM energy injections.

In the following sections, we consider possible new changes by taking
into account the capture of CHAMPs.
 
%%%%%%%%%%%%%%%%%%%%%%%%%%%%%%%%%%%%%%%%%%%%%%%%%%%%%%%%%%%%%%%%%%%%%%
\subsubsection{Are there corrections on the evaluation for 
the primary energy injection by CHAMP decays ?}
%%%%%%%%%%%%%%%%%%%%%%%%%%%%%%%%%%%%%%%%%%%%%%%%%%%%%%%%%%%%%%%%%%%%%%
Since the binding energy of CHAMP bound states is below the nuclear
binding energy of the light elements, the recoil of nucleons inside
the captured light element due to CHAMP decays would not destroy the
light element.  On the other hand, the decay products of the bound
CHAMP might directly hit  the bound light element and destroy it.
Let us consider the case that the primary decay product is a charged
lepton as an example. Of course,  if the lepton is a tau, the tau lepton
soon decays into hadronic particles.  However, the lifetime of a tau
lepton is long enough to go through  the Bohr radius of
the bound state.  Thus we will deal with all kinds of leptons 
in a similar fashion.

Naively we may speculate that the light elements are distributed inside the
radius $r_X^{-1}\simeq A^{-1/3}m_{\pi}$, and the CHAMP stays
somewhere inside the radius. Then the number density of quarks (or nucleons/
nuclei) inside a bound nucleus is roughly, 
\begin{eqnarray}
n_{\text{bound}X}\sim A(\frac{1}{r_X})^3
\simeq  m_{\pi}^3. 
\end{eqnarray}
The mean free path is roughly estimated by
\begin{eqnarray}
\lambda_{\text{mfp}}\simeq
\frac{1}{\sigma\times n_{\text{bound}X}},
\end{eqnarray}
where we have chosen $\sigma\simeq 2\pi \alpha^2/t$ where $t$ is the
Mandelstam variable $t$ for the momentum transfer from a primary decay
product (a charged lepton)  to a bound light element.  Then, the naive
probability of the primary decay product (a charged lepton) 
scattering off a quark (or nucleon/nucleus) inside the nucleus is
\begin{eqnarray}
\text{Prob}\simeq
\frac{2r_X}{\lambda_{\text{mfp}}}
\sim O(10^{-9})\left[ \frac{(10\text{GeV})^2}{t} \right]
\left[ \frac{A^{1/3}}{2} \right].
\end{eqnarray}

Among light elements, $^4$He destruction would be most dangerous. If 
we assume that all of $^4$He's are completely captured by CHAMPs,  if
such a probability is below $10^{-4}$, the change on D/H, $^3$He/H
abundances due to  the direct collision will be below
the $O(10^{-5})$ level,  which may not disagree  with observed abundances.
Elements heavier than the destroyed parent nuclei should not
be directly produced significantly.  \footnote{However, in a recent
work~\cite{Kaplinghat:2006qr}, they have pointed out the possibility
that energetic T and $^{3}$He which are produced from the destruction
of the bound $^4$He  can nonthermally produce sizable amount of
$^6$Li.}

We can find that, if the momentum transfer from primary decay-product
is hard ($t>(100\text{MeV})^2$)\footnote{The energy transfer due to
the momentum transfer $< (100$MeV)$^2$ may be below the typical
threshold $\sim \cal{O}$(10) MeV to destroy a bound light element by
NR nucleon/nuclei  scattering inside the light element.}, the light
element bound by a CHAMP is sufficiently transparent  and may not
disturb the SBBN prediction for elements lighter than $^4$He.  In this
case, for the evaluation of the primary energy injection,  past
studies in the literature  will be a good approximation, which considered
that CHAMPs are freely propagating in a thermal bath.  For the secondary
products through the hadronization of a recoiled quark or direct
production of nucleons/nuclei,  the above probability will be identified
as the hadronic branching ratio for  a CHAMP decay, which may provide
only negligible effects on elements lighter than $^4$He. On the other
hand, we can consider another extreme case where the momentum transfer
is sufficiently soft.   For example, if the energy is smaller than the
nuclear binding energy,  the charged lepton of decay products could
not inelastically scatter off the bound light element.  In the middle
range between them, we may have to simultaneously consider the direct
collision and EM/Hadronic cascade induced by the late decay which was
considered before.

In the case of hadronic decays, we may replace $\alpha$ by the  strong
coupling $\alpha_s$ in the above estimation.  Then, we find that for a
sufficiently hard momentum transfer $t>(1\text{GeV})^2$, the bound
light element  is still transparent.

%%%%%%%%%%%%%%%%%%%%%%%%%%%%%%%%%%%%%%%%%%%%%%%%%%%%%%%%%%%%%%%%%%%%%%
\subsubsection{Other new possible corrections to BBN with late time
energy injection} 
%%%%%%%%%%%%%%%%%%%%%%%%%%%%%%%%%%%%%%%%%%%%%%%%%%%%%%%%%%%%%%%%%%%%%%
There are three types of other possible effects on light element
abundances by the late-time decaying CHAMPs when some fraction of such
light elements are captured by CHAMPs.

The first type originates from the change on the Coulomb barrier and
the kinematics of background light elements  as targets for nonthermal
processes  by their own bound states.  This change could be important
for the hadronic-decay scenario.  The injected high-energy hadrons
eventually lose their energy due to the thermal interactions and
become nonrelativistic and collide with the background light
elements. If the target nuclei are captured by CHAMPs, the reaction
rates for various hadronic processes might be different from
experimental values.

The second type is related to the change of the {\it Q} value.  This might be
important for both high-energy hadronic and EM energy injections,
especially in high-energy photon injections. The injected high-energy
photons produce many soft photons through the EM cascade before the
scattering off background light elements. Then the spectrum of the
soft photons has a cutoff at the energy above  the threshold of
electron-positron pair creation, which depends on the cosmic time or
the cosmic temperature.  Only when the cutoff energy is higher than
the threshold energy of the photodissociation are the target nuclei 
destroyed.  The change in the {\it Q} value may modify the epoch when the
light element is destroyed by the photodissociation processes.
\footnote{On the other hand, we may also have to take care of the
destruction  of bound state due to huge soft photons from high energy
photon injection.}

The third type is related to neutron injection from late decays.  If
the decay occurs while some of the SBBN processes are still active, the high-energy 
hadronic injections might produce many neutrons. At late time
(t $>$ 100 sec), since neutrons have an extremely low abundance due to the
$\beta$ decay, the produced neutrons can significantly affect the light
element abundances because the related  nuclear reactions do not have a
Coulomb suppression. The neutron injection at around $10^3$ sec was
discussed as a solution to obtain low $^7$Be abundance by the
destruction  of SBBN $^7$Be through $^7$Be(n,p)$^7$Li and subsequently
$^7$Li(p,$\alpha$)$^4$He~\cite{Jedamzik:2004er,Kawasaki:2004qu,Kohri:2005wn}.
In our scenario, there may exist some differences from the previous
studies.  As we mentioned before, in CBBN, $^7$Be(n,$\alpha$)$^4$He
could be more important for the $^7$Be abundance than
$^7$Be(n,p)$^7$Li  because the center-of-mass energy in the process
can be completely different  from that of SBBN. Thus the neutron
injections from the late  decaying CHAMPs may enhance the destruction
ability of $^7$Be, and the effects could be different from the
noncaptured case.  Notice also that there may still exist unknown
errors even on the reaction rate of $^7$Be(n,p)$^7$Li  with captured
$^7$Be, as was discussed before.

The modification of the reaction rate of $^3$He(d,p)$^4$He might be
interesting with decaying CHAMP scenario below 1keV.  If a sizable
fraction of D is captured and the reaction rate of the process is enhanced 
due to the capture of a CHAMP, the destruction rate of $^3$He might become 
more rapid than the production due to late decays of CHAMPs, which may 
weaken the bound on $^3$He production due to the decay. This new
possibility may relax the $^3$He/D bound, which is generally the
most severe constraint on radiatively or hadronically decaying
massive-particle scenarios at $t>10^7$sec.

Considering the above possibilities, it is important to reanalyze
the effect of the late-time decaying CHAMPs in their bound states
with the light elements~\cite{takayama:2006}.

%%%%%%%%%%%%%%%%%%%%%%%%%%%%%%%%%%%%%%%%%%%%%%%%%%%%%%%%%%%%%%%%%%%%%%
\section{Discussions}
%%%%%%%%%%%%%%%%%%%%%%%%%%%%%%%%%%%%%%%%%%%%%%%%%%%%%%%%%%%%%%%%%%%%%%
If the CBBN prediction is not be significantly
disturbed  by late decays, the superWIMP dark matter scenario would be
interesting.  Here we discuss how much relic of CHAMPs might be
allowed in this scenario.  Since the number density of CHAMPs is
important to evaluate the capture rate  of light element, we consider
the possible constraints on the number  density of CHAMPs, assuming
that the  whole dark matter originates from the two body decay of
CHAMPs into a dark matter and a SM particle.  We consider the free
streaming by the whole  dark matter produced from CHAMP
decays. The relic density of CHAMP is
\begin{eqnarray}
\Omega_C=\frac{m_C}{m_{\text{DM}}}\Omega_{\text{DM}}
\end{eqnarray}
 As we found before, the capture rate is governed by the number density of 
the CHAMP.
\begin{eqnarray}
\frac{n_C}{n_{\gamma}}=\frac{n_{\text{DM}}}{n_{\gamma}}
=3\times 10^{-11}\frac{100\text{GeV}}{m_{\text{DM}}}
\frac{\Omega_{\text{DM}}}{0.23}
\end{eqnarray}
Hence the lighter mass of the dark-matter allows larger CHAMP
abundance.  Since keV warm dark matter is still allowed from
Ly$\alpha$ data ~\cite{Viel:2005qj}, we naively require that the dark
matter is nonrelativistic at $T=$keV.  Then we find a following
condition.
\begin{eqnarray}
u<1.0 \sqrt{\frac{10^6\text{sec}}{t}}
\end{eqnarray}
where $u=\sqrt{|p^ip_i|}/m$ and $p_i$ is the three momentum of dark
matter.  Assuming the two body decay, the four velocity at the decay time
is $u=(m_{\text{CHAMP}}^2-m_{\text{DM}}^2)/2m_{\text{DM}}m_{\text{CHAMP}}$.
Then we find that, for lifetime $\sim 10^4$sec, $u\sim 20$ may be
allowed. Then it is  possible to take $n_{\text{CHAMP}}/n_{\gamma}
\sim O(10^{-9})$, which will lead to a considerable capture rate.

For the case that the decaying CHAMPs  contribute only to part of
the dark matter, or their contribution is negligible,  the
above constraint may not be  applicable.

%%%%%%%%%%%%%%%%%%%%%%%%%%%%%%%%%%%%%%%%%%%%%%%%%%%%%%%%%%%%%%%%%%%%%%
\section{Conclusion}
%%%%%%%%%%%%%%%%%%%%%%%%%%%%%%%%%%%%%%%%%%%%%%%%%%%%%%%%%%%%%%%%%%%%%%
In this paper, we have discussed the role of long-lived charged
particle  during/after the BBN epoch.  We found that the existence of
CHAMP during the BBN epoch can change the light element abundances if
the capture rate of CHAMP by light elements is sufficiently large.  Since
the bound state for heavier elements tends to be more stable against
the destruction by the  background photon,  the abundances are
modified only for heavier elements such as  Li and Be, thanks to the
capture at an earlier time before the nuclear reactions decouple. On
the other hand, the abundances of lighter elements such as D, T,
$^3$He, and $^4$He are unchanged.  In fact, even though more work needs to be done 
to find quantitative results, we have shown
that the capture of CHAMPs may possibly  have some impact on the BBN
prediction of the primordial $^7$Li abundance.  Our approach to
consider the cosmological effects of the formation of 
the CHAMP bound states should also be attractive in some particle physics
models~\cite{Gluino,DoubleCharge}.

To understand CBBN more correctly, we need to understand  the nuclear
fusion rates and the capture rates more precisely. However, unfortunately 
there are still some uncertainties in the
experimental data of the reaction rates at present. We expect that the
future nuclear experiments will clarify these points. If future
collider experiments find  a signal of long-lived charged particle
inside the detector, the measurement  of lifetime and decay properties
of the charged particle will provide new insights to understand the
phenomena in the early universe in turn.

%%%%%%%%%%%%%%%%%%%%%%%%%%%%%%%%%%%%%%%%%%%%%%%%%%%%%%
\section{Acknowledgment}
%%%%%%%%%%%%%%%%%%%%%%%%%%%%%%%%%%%%%%%%%%%%%%%%%%%%%%
F.T would like to thank Bryan T. Smith, Jonathan L. Feng,  Jose
A.R. Cembranos for discussions in the early stage of this work and
Maxim Pospelov, Manoj Kaplinghat, Arvind Rajaraman for discussions in
SUSY06  and Csaba Csaki, Toichiro Kinoshita, Maxim Perelstein for valuable
suggestions. K.K was supported in part by NASA grant NNG04GL38G. 
F.T is supported in part by the NSF grant PHY-0355005.

%%%%%%%%%%%%%%%%%%%%%%%%%%%%%%%%%%%%%%%%%%%%%%%%%%%%%%

\end{document}